\documentclass[
  aps,
  prl,
  reprint,
  superscriptaddress,
  amsfonts,amssymb,amsmath
]{revtex4-2}

\usepackage{newtxtext}
\usepackage{bm}
\usepackage{latexsym}
\usepackage{amsthm}
\usepackage{mathtools}
\usepackage{xcolor}
\usepackage{physics}
\usepackage{graphicx}
\usepackage{float}
\usepackage{algorithm}
\usepackage{algpseudocode}
\usepackage{hyperref}
\hypersetup{
  setpagesize  = false,
  colorlinks   = true,
  urlcolor     = blue,
  linkcolor    = blue,
  citecolor    = blue
}
\usepackage{tikz}
\usetikzlibrary{arrows.meta,decorations.pathreplacing}

\theoremstyle{plain}

\newcommand{\beginendmatt}{
  \twocolumngrid
  \setcounter{equation}{0}
  \renewcommand{\theequation}{A\arabic{equation}}%
  \renewcommand{\theHequation}{A.\arabic{equation}}%
  \setcounter{figure}{0}
  \renewcommand{\thefigure}{A\arabic{figure}}%
  \renewcommand{\theHfigure}{A.\arabic{figure}}%
  \setcounter{table}{0}
  \renewcommand{\thetable}{A\arabic{table}}%
  \renewcommand{\theHtable}{A.\arabic{table}}%
  \setcounter{section}{0}
  \renewcommand{\thesection}{A\Roman{section}}%
  \renewcommand{\theHsection}{A.\Roman{section}}%
  \makeatletter
}

\newcommand{\beginsupplement}{
  \setcounter{equation}{0}
  \renewcommand{\theequation}{S\arabic{equation}}%
  \renewcommand{\theHequation}{S.\arabic{equation}}%
  \setcounter{figure}{0}
  \renewcommand{\thefigure}{S\arabic{figure}}%
  \renewcommand{\theHfigure}{S.\arabic{figure}}%
  \setcounter{table}{0}
  \renewcommand{\thetable}{S\arabic{table}}%
  \renewcommand{\theHtable}{S.\arabic{table}}%
  \setcounter{section}{0}
  \renewcommand{\thesection}{S\Roman{section}}%
  \renewcommand{\theHsection}{S.\Roman{section}}%
  \setcounter{secnumdepth}{2}% prl option disables section numbers
  \makeatletter
}

\begin{document}

%%% title %%%
\title{Efficient Sampling for Many-Body Fermionic Non-Gaussianity}

%%% authors %%%
\author{Ryota Matsuda}
\thanks{These authors contributed equally to this work.}
\affiliation{Department of Physics, The University of Tokyo, 7-3-1 Hongo, Bunkyo-ku, Tokyo 113-0033, Japan}

\author{Masahiro Hoshino}
\thanks{These authors contributed equally to this work.}
\affiliation{Department of Physics, The University of Tokyo, 7-3-1 Hongo, Bunkyo-ku, Tokyo 113-0033, Japan}

\author{Yuto Ashida}
\affiliation{Department of Physics, The University of Tokyo, 7-3-1 Hongo, Bunkyo-ku, Tokyo 113-0033, Japan}
\affiliation{Institute for Physics of Intelligence, The University of Tokyo, 7-3-1 Hongo, Bunkyo-ku, Tokyo 113-0033, Japan}

%%% date %%% 
\date{\today}

%%% abstract %%%
\begin{abstract}
  Fermionic non-Gaussianity is a key resource for universal quantum computation, and its behavior in quantum many-body systems has attracted growing interest.
  Recently, a convolution-based measure, which we call the magic R\'enyi entropy (MRE), has been proposed as a resource measure of non-Gaussianity, but its evaluation is limited to small systems due to the  computational cost growing exponentially with system size.
  In this work, we develop a perfect-sampling method to calculate the second-order MRE from matrix product states (MPSs).
  Our method uses a bounded estimator to control sampling fluctuations and a recursive sweep to draw the samples directly from the input MPS, which allows us to generate each sample with $\mathcal O(nD^3)$ time, where $n$ and $D$ denote the number of modes and MPS bond dimension, respectively.
  We benchmark our method on the XXZ chain with up to $128$ sites, demonstrating that the MRE captures intricate higher-order correlations at large scales, which are inaccessible to covariance-based measures.
  These results provide a computational tool that quantitatively evaluates many-body fermionic non-Gaussianity by accounting for higher-order correlations.
\end{abstract}

\maketitle

%%%========== main text ==========%%%

%%%========== Introduction ==========%%%
Fermionic Gaussian states and dynamics are completely characterized by their two-point correlations (covariance matrix), making them efficiently simulable on classical computers~\cite{valiant2002quantum,knill2001fermionic,terhal2002classical,bravyi2004lagrangian}.
Consequently, non-Gaussian states or operations are essential resources for universal quantum computation~\cite{hebenstreit2019all}.
This resource-theoretic structure closely parallels the distinction between efficiently simulable Clifford circuits and nonstabilizer (magic) resources in spin systems~\cite{gottesman1998heisenberg,bravyi2005universal}.
In interacting electronic systems, fermionic non-Gaussianity also quantifies correlations beyond Gaussian or mean-field descriptions~\cite{bach1994generalized} and may provide a way to assess the limitations of Gaussian variational states~\cite{bravyi2017complexity,ashida2018solving,bravyi2019approximation,shi2020ultrafast,wang2020zerotemperature,boutin2021quantum,snyman2021efficient}.
The same perspective applies to quantum chemistry, where simulating correlated molecular electronic states beyond Hartree-Fock approximations is a major goal of quantum computation~\cite{mcardle2020quantum,bauer2020quantum}.

Recently, a resource measure based on the convolution operation~\cite{lyu2024fermionic,coffman2025measuring}, which we call the magic R\'enyi entropy (MRE)~\cite{matsuda2026quantum}, has been proposed as a valid quantifier of many-body fermionic non-Gaussianity that satisfies resource-theoretic axioms.
Specifically, the MRE with R\'enyi index $2$ was shown to be monotone under pure-to-pure fermionic Gaussian protocols that involve measurements and feedforward~\cite{matsuda2026quantum}.
The convolution-based framework also equally applies to bosonic~\cite{bu2025efficient,hahn2025measuring} and qubit/qudit systems, reducing to the stabilizer R\'enyi entropy (SRE)~\cite{leone2022stabilizer} in the latter case~\cite{bu2025stabilizer,matsuda2026quantum}.
Since the MRE is defined by the purity of a convolved state, it can be experimentally measured, e.g., via the SWAP test~\cite{ekert2002direct} or the interference of two copies~\cite{daley2012measuring,pichler2013thermal,islam2015measuring}.
While previously proposed covariance-matrix-based measures~\cite{sierant2026fermionic,tarabunga2026computable,leone2026fermionic,turkeshi2026williamson} provide computationally efficient diagnostics of fermionic non-Gaussianity, the MRE captures higher-order correlations~\cite{hoshino2026universal} that play an essential role in strongly correlated systems.
Since the number of these higher-order correlations grows exponentially with the system size, however, exact evaluation of the MRE has been limited to small system sizes.

Using matrix product states (MPSs) is a natural step for evaluating the MRE at much larger scales.
For the SRE in spin systems, several practical algorithms are already available, including  perfect Pauli sampling~\cite{haug2023stabilizer,lami2023nonstabilizerness}, replicated MPS representations~\cite{haug2023quantifying,tarabunga2024nonstabilizerness}, and Pauli-Markov chain sampling~\cite{tarabunga2023manybody}, among others~\cite{ding2025evaluating,collura2026nonstabilizerness,xiao2026exponentially,huang2026fast}.
These methods rely on the fact that the SRE can be written as the higher-order moments of the probabilistic distribution (participation entropy) in the Pauli string.
While the fermionic MRE inherits a similar structure in that it can be represented using higher-order moments of the characteristic function, its Grassmann nature requires careful treatment of signs, preventing it from being regarded as a participation entropy in the Majorana basis.
It is thus not obvious whether the fermionic MRE admits an efficient MPS algorithm analogous to those developed for the SRE.

In this Letter, we develop a perfect-sampling method for evaluating the second-order fermionic MRE from MPSs.
We introduce a bounded estimator that suppresses sampling fluctuations and a recursive contraction scheme that generates each sample directly from the input MPS.
These two ideas avoid the explicit construction of the beam-splitter output MPS and keep the sampling cost to a low-order polynomial in both the system size and MPS bond dimension.
We apply the method to an XXZ chain of spinful fermions with up to $128$ sites.
At the ferromagnetic point, the sampled MRE agrees with an exact result for every system size, and away from this point it resolves a saturation of the MRE that appears only at sizes beyond the reach of exact evaluation.

%%%========== Fermionic non-Gaussianity and magic R\'enyi entropy ==========%%%
\textit{Fermionic non-Gaussianity and magic R\'enyi entropy.}---
We first briefly review the resource theory of fermionic non-Gaussianity and the MRE as a resource measure.
For a system of $n$ fermionic modes, the annihilation and creation operators $c_j,c_j^\dagger$ satisfy the canonical anticommutation relations: $\{c_i,c_j^\dagger\}=\delta_{ij}$ and $\{c_i,c_j\}=\{c_i^\dagger,c_j^\dagger\}=0$.
It is convenient to introduce $2n$ Majorana operators $\gamma_{2j-1}=c_j+c_j^\dagger,\gamma_{2j}=i(c_j-c_j^\dagger)$, which satisfy $\{\gamma_\mu,\gamma_\nu\}=2\delta_{\mu\nu}$.
Writing $\bm{\gamma}=(\gamma_1,\gamma_2,\ldots,\gamma_{2n})^{\mathsf T}$, a fermionic Gaussian unitary is generated by a quadratic Majorana operator,
\begin{equation}
  U_G
  =\exp\!\left(\frac{1}{4}\bm{\gamma}^{\mathsf T}K\bm{\gamma}\right),
\end{equation}
where $K\in\mathbb{R}^{2n\times2n}$ and $K^{\mathsf T}=-K$.
The unitary acts linearly on the Majorana operators as $U_G\bm{\gamma}U_G^\dagger=O\bm{\gamma}$, where $O=e^{-K}\in\mathrm{SO}(2n)$.
The set of even-parity pure Gaussian states is defined as the orbit of the vacuum under the group $\mathcal G_n$ of Gaussian unitaries: $\mathrm{Gauss}^{(+)}_n=\left\{U_G\ket{0}^{\otimes n}\mid U_G\in\mathcal G_n\right\}$; the odd-parity sector is obtained by choosing any fixed odd-parity Gaussian reference state.

The MRE quantifies fermionic non-Gaussianity through the purity loss generated by a convolution operation.
For the second-order MRE, which is the focus of this work, we use the convolution implemented by the following balanced beam splitter unitary acting on two copies of the system labeled $A$ and $B$:
\begin{equation}
  U_{\mathrm{BS}}
  = \exp\!\left[\frac{\pi}{4}\sum_{j=1}^{n}\left(c_{A,j}^\dagger c_{B,j}-c_{B,j}^\dagger c_{A,j}\right)\right].
  \label{eq:beam_splitter}
\end{equation}
For a pure input state $\rho=\ketbra{\psi}{\psi}$, the convolved state is $\mathcal{C}(\rho) = \Tr_B|\Phi\rangle\langle\Phi|$, where $\ket{\Phi}=U_{\mathrm{BS}}\ket{\psi}_A\ket{\psi}_B$.
The two outputs remain separable and $\mathcal{C}(\rho)$ is pure if and only if the input state is Gaussian. Thus, any loss of purity reflects non-Gaussianity of the input.
Accordingly, the second-order MRE is defined from the purity $\nu_2=\Tr[\mathcal{C}(\rho)^2]$ as $M_2(\psi)=-\ln\nu_2\geq 0$.
For pure fermionic states, the second-order MRE satisfies the following properties~\cite{matsuda2026quantum}:
\begin{itemize}
  \item \textit{Faithfulness.}  $M_2(\psi)=0\Longleftrightarrow\ket{\psi}\text{ is Gaussian}$.
  \item \textit{Invariance under Gaussian unitaries.} $M_2(U_G\ket{\psi})=M_2(\psi)$ for every $U_G\in\mathcal G_n$.
  \item \textit{Additivity.} $M_2(\psi\otimes\phi)=M_2(\psi)+M_2(\phi)$.
  \item \textit{Monotonicity.} A fermionic Gaussian protocol may combine Gaussian unitaries, pure Gaussian state ancillas, occupation-number measurements, partial traces, classical randomness, and feedforward.
        If such a protocol $\mathcal E$ deterministically maps a pure input to a pure output, $\mathcal E(\ketbra{\psi}{\psi})=\ketbra{\varphi}{\varphi}$, then $M_2(\varphi)\leq M_2(\psi)$.
\end{itemize}
Moreover, the linear MRE $M_2^{\mathrm{lin}}=1-\nu_2$ satisfies strong monotonicity under fermionic Gaussian protocols, meaning that if a protocol maps $\ket{\psi}$ to pure outputs $\ket{\varphi_\lambda}$ with probabilities $p_\lambda$, then $\sum_\lambda p_\lambda M_2^{\mathrm{lin}}(\varphi_\lambda)\leq M_2^{\mathrm{lin}}(\psi)$.
Together, these properties establish $M_2$ as a valid resource measure of fermionic non-Gaussianity.

In general, however, computing the second-order MRE directly from a many-body wave function is prohibitively expensive; a direct evaluation in the occupation basis involves exponentially many configurations.
An explicit replica-MPS calculation is also costly because the beam splitter enlarges the virtual bond dimension before taking the contraction to compute the purity.
For an input MPS of bond dimension $D$, the purity contracts four copies of the beam-splitter output through a transfer matrix of dimension $\mathcal O(D^8)$.
These bottlenecks motivate a sampling method that avoids constructing the replicated tensor network.

%%%========== A bounded estimator of the purity ==========%%%
\textit{Bounded estimator of the purity.}---
To this end, we recast the purity $\nu_2$ as an expectation value that can be estimated efficiently from Fock-basis samples of the output.
Write $\Phi(\bm a,\bm b)=\braket{\bm a,\bm b}{\Phi}$, where $\bm a,\bm b\in\{0,1\}^n$ label the Fock configurations of outputs $A$ and $B$.
Tracing out $B$ gives
\begin{equation}
  \mathcal{C}(\rho)(\bm a,\bm a')
  =
  \sum_{\bm b}
  \Phi(\bm a,\bm b)\,
  \overline{\Phi(\bm a',\bm b)},
\end{equation}
where the overline denotes complex conjugation.
This leads to
\begin{align}
  \nu_2
  =
  \sum_{\bm a,\bm a',\bm b,\bm b'}
  \overline{\Phi(\bm a,\bm b)}\,
  \overline{\Phi(\bm a',\bm b')}\,
  \Phi(\bm a',\bm b)\,
  \Phi(\bm a,\bm b').\label{nu2}
\end{align}
Let $X=((\bm a,\bm b),(\bm a',\bm b'))$ denote a pair of independent Fock configurations drawn from the product of the Born distribution $q(X)=|\Phi(\bm a,\bm b)|^2|\Phi(\bm a',\bm b')|^2$.
On the support of $q$, we define the amplitude ratio associated with exchanging the $A$-output configurations by
\begin{equation}
  r(X)
  =
  \frac{
    \Phi(\bm a',\bm b)\,
    \Phi(\bm a,\bm b')
  }{
    \Phi(\bm a,\bm b)\,
    \Phi(\bm a',\bm b')
  }.
\end{equation}
Because $q(X)r(X)$ reproduces the corresponding summand in Eq.~\eqref{nu2}, the purity is given by $\nu_2=\mathbb E_q[r]$.
An estimator of the same form as $r(X)$ is used in variational Monte Carlo simulations to evaluate the purity of a reduced density matrix~\cite{zhang2011entanglement}.

Since the unbiased estimator $r(X)$ is unbounded, we use the Rao-Blackwellization to construct a bounded unbiased estimator.
Specifically, for each sample $X$, we define its exchanged partner $TX$ by swapping $\bm a$ and $\bm a'$ while leaving $\bm b$ and $\bm b'$ unchanged:
\begin{equation}
  TX
  =
  ((\bm a',\bm b),(\bm a,\bm b')).
\end{equation}
When $r(X)\neq0$, the two configurations satisfy $q(TX)=q(X)|r(X)|^2$ and $r(TX)=1/r(X)$.
Averaging $r$ over each exchange pair with the corresponding conditional weights gives
\begin{equation}
  h(X)
  =
  \frac{q(X)r(X)+q(TX)r(TX)}
  {q(X)+q(TX)}
  =
  \frac{2\operatorname{Re}r(X)}{1+|r(X)|^2},
\end{equation}
where $h(X)=0$ when $r(X)=0$, while a fixed point $TX=X$ has $h(X)=r(X)=1$.
We note that this conditional expectation preserves the mean while reducing the variance, i.e., $\mathbb E_q[h]=\mathbb E_q[\operatorname{Re}r]=\nu_2$ and $\operatorname{Var}_q(h)\leq\operatorname{Var}_q(\operatorname{Re}r)$~\cite{rao1945information,blackwell1947conditional}.
Moreover, the relation $2|\operatorname{Re}r|\leq1+|r|^2$ implies $-1\leq h\leq1$, eliminating the large fluctuations caused by small denominators in $r$.

Evaluating the bounded estimator requires two Born samples and two additional amplitude evaluations for the exchanged configurations.
The sampled configurations provide $(\bm a,\bm b)$ and $(\bm a',\bm b')$, while the ratio additionally requires $\Phi(\bm a',\bm b)$ and $\Phi(\bm a,\bm b')$.
Thus, the remaining task is to sample exactly from $|\Phi|^2$ and evaluate prescribed amplitudes of $\ket{\Phi}$ without explicitly constructing the replicated MPS.

%%%========== Perfect sampling from an MPS ==========%%%
\textit{Perfect sampling from an MPS.}---
To sample and evaluate the amplitudes efficiently, we represent the input state as a normalized right-canonical MPS with one site per mode:
\begin{equation}
  \ket*{\psi} = \sum_{\bm{x}\in\{0,1\}^n} \psi(\bm x)\ket{\bm x},\quad \psi(\bm x) = A_1^{x_1}A_2^{x_2}\cdots A_n^{x_n}.
\end{equation}
The tensors satisfy $A_j^x\in\mathbb C^{D_{j-1}\times D_j}$ and $D_0=D_n=1$, while $\ket{\bm x}=(c_1^\dagger)^{x_1}(c_2^\dagger)^{x_2}\cdots(c_n^\dagger)^{x_n}\ket{0}$ and $D=\max_jD_j$.
We work with a conventional MPS representation in this fixed Fock basis rather than a graded fermionic MPS~\cite{bultinck2017fermionic}.
We assume a definite fermion parity $P\in\{0,1\}$ such that $\psi(\bm x)=0$ unless $(\sum_jx_j)\bmod2=P$.
The tensors also satisfy $\sum_{x=0}^1A_j^x(A_j^x)^\dagger=I_{D_{j-1}}$.
Let $x_j,y_j$ denote the input occupations of copies $A,B$ and $a_j,b_j$ the corresponding output occupations after the convolution unitary, which we encode by $z_j=2a_j+b_j\in\{0,1,2,3\}$.

To sample $X$ from the probability distribution $q(X)$, we independently draw two output configurations from the Born distribution using a sequence of one-site conditional probabilities.
Let $P_\Phi(\bm z)=|\Phi(\bm z)|^2$ denote the Born probability of the output configuration $\bm z=(z_1,z_2,\ldots,z_n)$.
For a prefix $\bm z_{\leq j}=(z_1,z_2,\ldots,z_j)$, define its marginal probability by
\begin{equation}
  W_j(\bm z_{\leq j}) = \sum_{\bm z_{>j}} P_\Phi(\bm z),\quad W_0=1.
\end{equation}
Writing $(\bm z_{<j},z)$ for the prefix obtained by appending $z$ to $\bm z_{<j}$, the chain rule gives $P_\Phi(\bm z) = \prod_{j=1}^n w_j(z_j\,|\,\bm z_{<j})$, where the conditional probability of $z_j$ given the preceding prefix is
\begin{equation}
  w_j(z\,|\,\bm z_{<j}) = \frac{W_j(\bm z_{<j},z)}{W_{j-1}(\bm z_{<j})}.
\end{equation}
Sequentially drawing each $z_j$ from $w_j(z\,|\,\bm z_{<j})$ produces an exact sample from $P_\Phi$~\cite{ferris2012perfect}.
The recursion below evaluates these conditional probabilities and the prescribed amplitudes without enumerating exponentially many configurations.

The exact output amplitude can be evaluated recursively using two parity-resolved boundary matrices.
To see this, we first note that the amplitude can be written as follows, where the sum runs over $\bm x,\bm y\in\{0,1\}^n$:
\begin{equation}
  \Phi(\bm z)
  =
  s(\bm a,\bm b)
  \sum_{\bm x,\bm y}
  \left[
    \prod_{j=1}^n
    u_{z_j,2x_j+y_j}(-1)^{x_jp_{j-1}}
    \right]
  \psi(\bm x)\psi(\bm y).
\end{equation}
Here, $u_{z,2x+y}=\langle a,b|U_{\mathrm{BS}}^{(j)}|x,y\rangle$, with rows labeling outputs and columns labeling inputs, $p_j=(\sum_{k=1}^jy_k)\bmod2$ with $p_0=0$ being the parity of the first $j$ sites of input copy $B$, and the factor $(-1)^{x_jp_{j-1}}$ accounts for moving $x_j$ past the preceding occupations of copy $B$ in the sitewise contraction.
Reordering the output modes from $(A_1,B_1,A_2,B_2,\ldots,A_n,B_n)$ to $(A_1,A_2,\ldots,A_n,B_1,B_2,\ldots,B_n)$ contributes the sign
\begin{equation}
  s(\bm a,\bm b)
  =
  (-1)^{\sum_{j=1}^na_j\sum_{k<j}b_k}.
\end{equation}
Starting from $\mathcal B_0^{(0)}=(1)$ and $\mathcal B_0^{(1)}=(0)$, we define two boundary matrices for each fixed prefix $\bm z_{\leq j}$ by
\begin{equation}\label{eq:boundary_recursion}
  \begin{aligned}
    \mathcal B_j^{(p_j)}(\bm z_{\leq j})
     & =
    \sum_{\substack{p_{j-1},x,y\in\{0,1\}\\p_{j-1}\oplus y=p_j}}
    u_{z_j,2x+y}(-1)^{xp_{j-1}}
    \\
     & \quad\times
            (A_j^x)^{\mathsf T}
    \mathcal B_{j-1}^{(p_{j-1})}(\bm z_{<j})
    A_j^y.
  \end{aligned}
\end{equation}
Since the fermionic sign at each site depends on the preceding occupations of copy $B$ only through their parity $p_{j-1}$, the recursion retains only the updated bit $p_j=p_{j-1}\oplus y$.
At $j=n$, the boundary matrix is a scalar and the fixed input parity leaves only $p_n=P$, giving $\Phi(\bm z)=s(\bm a,\bm b)\mathcal B_n^{(P)}(\bm z)$.

\begin{figure}[tb]
  \centering
  \includegraphics[width=\columnwidth]{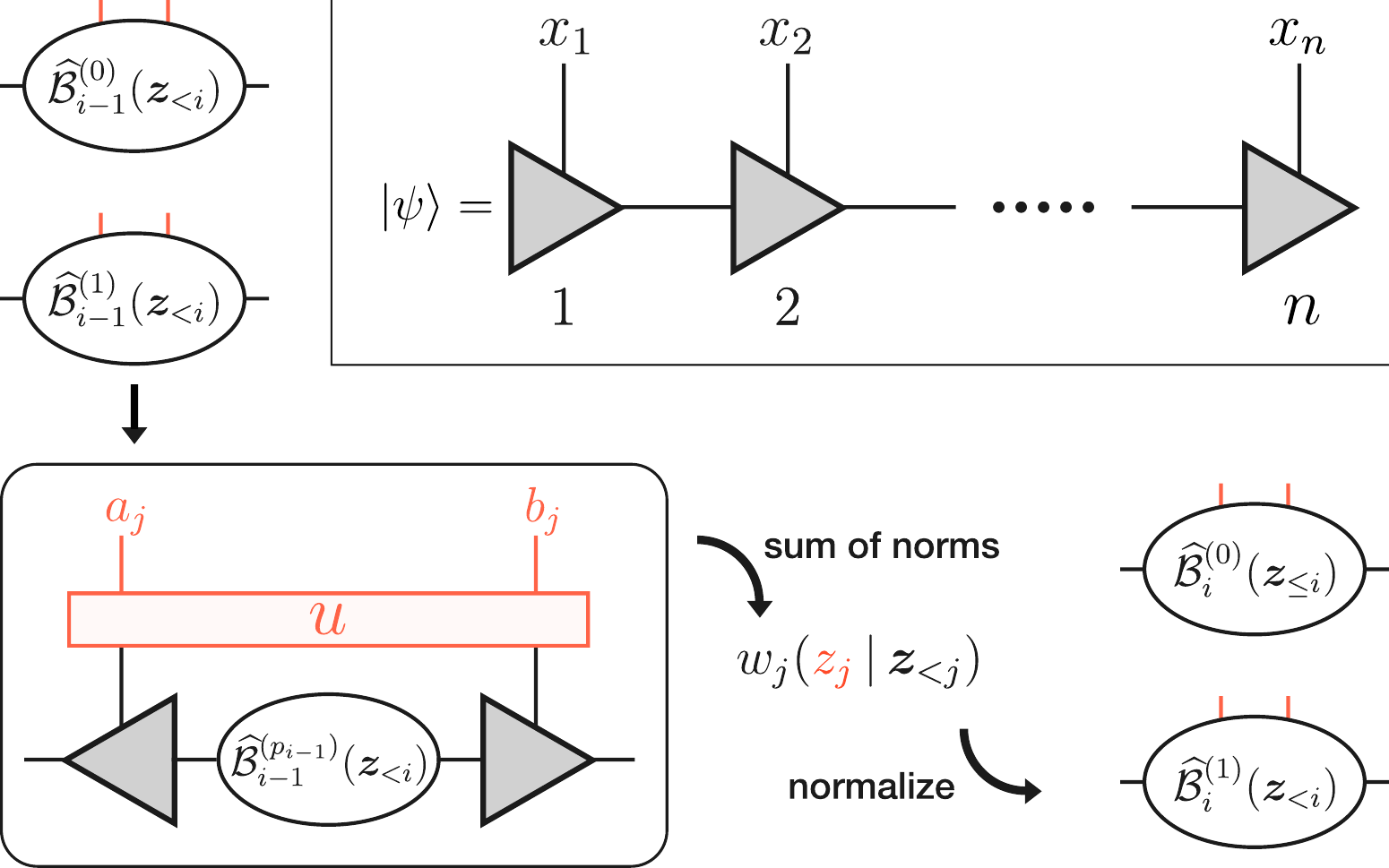}
  \caption{Schematic illustration of the perfect-sampling sweep.
    The two parity-resolved boundary matrices encode the sampled prefix and are contracted with the local tensors of the input MPS and the beam splitter to obtain the conditional probabilities at each site.
    After the local output is sampled, the corresponding matrices are normalized and passed to the next site, yielding an exact Born sample without constructing the convolved MPS.}
  \label{fig:perfect_sampling_sweep}
\end{figure}

As illustrated in Fig.~\ref{fig:perfect_sampling_sweep}, the two current boundary matrices contain all information of the prefix needed to compute the next conditional probability.
For a right-canonical MPS, the marginal probability satisfies
\begin{equation}\label{eq:conditional_prob}
  W_j(\bm z_{\leq j})
  =
  \sum_{p_j=0}^1
  \left\|
  \mathcal B_j^{(p_j)}(\bm z_{\leq j})
  \right\|_{\mathrm F}^2,
\end{equation}
where $\|M\|_{\mathrm F}^2=\Tr(MM^\dagger)$ is the squared Frobenius norm; see the End Matter for the derivation.
Applying Eq.~\eqref{eq:boundary_recursion} to the current matrices for each $z\in\{0,1,2,3\}$ and using Eq.~\eqref{eq:conditional_prob} yields the four conditional probabilities without storing earlier boundaries or contracting the unobserved suffix.
For a selected prefix with nonzero probability, we apply the recursion to the normalized matrices $\widehat{\mathcal B}_{j-1}^{(p_{j-1})}\coloneqq\mathcal B_{j-1}^{(p_{j-1})}(\bm z_{<j})/\sqrt{W_{j-1}(\bm z_{<j})}$, which gives candidates whose squared Frobenius norms sum to $w_j(z\,|\,\bm z_{<j})$ for each $z$.
After drawing $z_j$ according to $w_j(z\,|\,\bm z_{<j})$, dividing the boundary matrices by $\sqrt{w_j(z_j\,|\,\bm z_{<j})}$ produces $\widehat{\mathcal B}_j^{(0,1)}$ for the next site.
To recover the amplitude without numerical underflow, we set $\lambda_0=0$ and update
\begin{equation}\label{eq:lambda_update}
  \lambda_j
  :=
  \frac12\log W_j(\bm z_{\leq j})
  =
  \lambda_{j-1}
  +
  \frac12\log w_j(z_j\,|\,\bm z_{<j}).
\end{equation}
Along the selected path, we have $\mathcal B_j^{(p_j)}=e^{\lambda_j}\widehat{\mathcal B}_j^{(p_j)}$, and the endpoint amplitude is
\begin{equation}\label{eq:normalized_amplitude}
  \Phi(\bm a,\bm b)
  = s(\bm a,\bm b) e^{\lambda_n} \widehat{\mathcal B}_n^{(P)}.
\end{equation}
This recursion provides both the Born samples and the prescribed amplitudes needed to estimate $\nu_2$.
The sampling sweep is summarized in Algorithm~\ref{alg:perfect_sampling}.
\begin{algorithm}[H]
  \caption{Perfect Born sampling from an MPS.}
  \label{alg:perfect_sampling}
  \begin{algorithmic}[1]
    \Require A right-canonical MPS with definite fermion parity $P$.
    \State Set $\widehat{\mathcal B}_0^{(0)}\gets(1)$, $\widehat{\mathcal B}_0^{(1)}\gets(0)$, and $\lambda_0\gets0$.
    \For{$j=1,2,\ldots,n$}
    \State \parbox[t]{\dimexpr\linewidth-\algorithmicindent-\algorithmicindent-2pt\relax}{\raggedright Compute $w_j(z\,|\,\bm z_{<j})$ for $z\,{=}\,0,1,2,3$ using Eqs.~\eqref{eq:boundary_recursion} and~\eqref{eq:conditional_prob}.\par}
    \State Draw $z_j$ according to $w_j(z\,|\,\bm z_{<j})$.
    \State \parbox[t]{\dimexpr\linewidth-\algorithmicindent-\algorithmicindent-2pt\relax}{\raggedright Update $\widehat{\mathcal B}_j^{(0,1)}$ by the normalization above and $\lambda_j$ using Eq.~\eqref{eq:lambda_update}.\par}
    \EndFor
    \State \Return $\bm z$ and its amplitude from Eq.~\eqref{eq:normalized_amplitude}.
  \end{algorithmic}
\end{algorithm}

Two calls to Algorithm~\ref{alg:perfect_sampling} generate $X=((\bm a,\bm b),(\bm a',\bm b'))\sim q$ and return the denominator amplitudes $   \Phi(\bm a,\bm b)\Phi(\bm a',\bm b')$ in $r(X)$.
The crossed amplitudes $\Phi(\bm a',\bm b)\,\Phi(\bm a,\bm b')$ are obtained by deterministic sweeps that fix $z_j=2a_j+b_j$ at every site.
If the norm of the boundary matrices becomes zero during a sweep, the corresponding amplitude is zero and the sweep can be terminated immediately.
Thus, one sample of $h(X)$ requires two sampling sweeps and at most two deterministic amplitude-evaluation sweeps.
Because the two-copy output has even total parity, unequal parities of $\sum_ja_j$ and $\sum_ja_j'$ imply $h(X)=0$, in which case the two deterministic sweeps need not be performed.
Repeating the procedure $\mathcal N$ times yields samples $X_1,X_2,\ldots,X_{\mathcal N}$ and the estimator $\widetilde{\nu}_2=\mathcal N^{-1}\sum_{\mu=1}^{\mathcal N}h(X_\mu)$.

%%%========== Statistical error and computational cost ==========%%%
\textit{Statistical error and computational cost.}---
We now discuss the statistical error and computational cost of the sampling method.
The unbiased estimator of the purity has standard error $s_h/\sqrt{\mathcal N}$, where $s_h^2$ is the sample variance of $h(X_\mu)$.
When $\widetilde{\nu}_2>0$, the corresponding estimator of the MRE is $\widetilde M_2=-\ln\widetilde{\nu}_2$, whose variance to leading order in $\mathcal N^{-1}$ is $\operatorname{Var}(\widetilde M_2)\simeq\operatorname{Var}_q(h)/(\mathcal N\nu_2^2)$.
Thus, to achieve a target standard error $\epsilon_M$, one needs $\mathcal N\simeq\operatorname{Var}_q(h)/(\epsilon_M^2\nu_2^2)$ samples.
Since $|h|\leq1$, this sample count is at most $e^{2M_2}/\epsilon_M^2$. While this cost can be exponentially large for an extensive $M_2$, numerical results suggest that the computation remains feasible at least up to $\sim10^{2}$ sites (see the Supplemental Material~\cite{supplemental}).

The total runtime follows by combining the required sample count with the cost of evaluating each sample.
At each site, the eight matrix contractions indexed by $p_{j-1},x,y$ are evaluated once and combined with $u_{z,2x+y}$ for the four values of $z$ to obtain all trial updates.
This contraction order avoids dense $D^2\times D^2$ transfer tensors, and each sampling or prescribed-amplitude sweep costs $\mathcal O(nD^3)$ time and $\mathcal O(D^2)$ additional working memory.
Since each value of $h$ requires only a constant number of sweeps, the full calculation costs $\mathcal O(\mathcal NnD^3)$ time with the same asymptotic working memory.
Note that they can be generated in parallel by processes with independent random streams, which reduces the wall-clock time accordingly.
When particle-number conservation is explicitly encoded in the input MPS~\cite{singh2011tensor}, the recursion can be restricted to the allowed particle-number sectors, substantially reducing the computational cost in practice while retaining the same asymptotic scaling (see End Matter for details).

%%%========== Application to an XXZ chain of spinful fermions ==========%%%
\textit{Numerical benchmark.}---
We demonstrate our method on an XXZ chain of spinful fermions.
We take fermions $c_{j\sigma}$ with $\sigma=\uparrow,\downarrow$ on an open chain of $L$ lattice sites, i.e., $n=2L$ modes, and restrict them to the subspace with one fermion per site, where the Hamiltonian is the spin-1/2 XXZ chain
\begin{equation}\label{eq:xxz_chain}
  H=\sum_{j=1}^{L-1}\left(S_j^xS_{j+1}^x+S_j^yS_{j+1}^y+\Delta S_j^zS_{j+1}^z\right)
\end{equation}
with $S_j^+=c_{j\uparrow}^\dagger c_{j\downarrow}$ and $S_j^z=(n_{j\uparrow}-n_{j\downarrow})/2$.
The Hamiltonian conserves $N_\uparrow$ and $N_\downarrow$ separately.
For even $L$, we consider its ground state at $N_\uparrow=N_\downarrow=L/2$, which is unique for every finite $L$~\cite{supplemental}.
Particle-number conservation and the spin-flip symmetry of this state fix $\ev{c_{i\sigma}^\dagger c_{j\sigma'}}$ to $\delta_{ij}\delta_{\sigma\sigma'}/2$, and the covariance matrix becomes identical for any $\Delta$~\cite{supplemental}.
Consequently, covariance-matrix-based measures~\cite{sierant2026fermionic,tarabunga2026computable,leone2026fermionic,turkeshi2026williamson} cannot distinguish this family of ground states.
For example, the fermionic antiflatness $\mathcal F_k$~\cite{sierant2026fermionic} takes its maximal value $2L$ for every $k$ and any $\Delta$.

\begin{figure}[tb]
  \centering
  \includegraphics[scale=0.75]{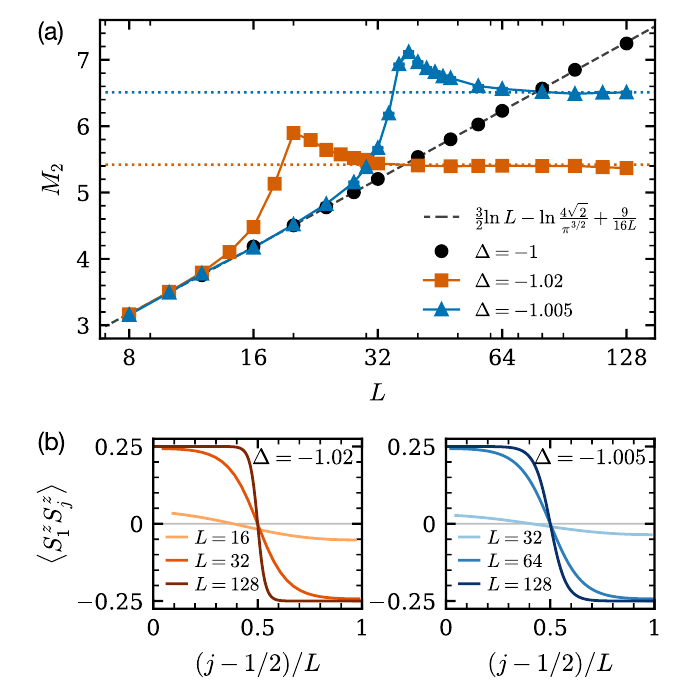}
  \caption{
    (a) Second-order MRE of the ground state as a function of the number of sites $L$.
    At $\Delta=-1$, the dashed black line is the large-$L$ asymptotic form~\eqref{eq:dicke_asymptotic}, and the circles are the perfect sampling from the exact MPS.
    The squares and triangles are the perfect sampling from the DMRG states at $\Delta=-1.02$ and $-1.005$, connected by solid curves as guides, and the dotted horizontal lines are Eq.~\eqref{eq:domain_wall_mre}.
    The standard errors are smaller than $0.01$ for all points.
    (b) shows the spin-spin correlation function $\ev{S_1^zS_j^z}$ at $\Delta=-1.02$ for $L=16,32,128$ and $\Delta=-1.005$ for $L=32,64,128$.}
  \label{fig:xxz_mre}
\end{figure}

At $\Delta=-1$, the Hamiltonian becomes the ferromagnetic Heisenberg model via a sublattice rotation, whose ground state is the spin-$L/2$ Dicke state with $L/2$ up spins~\cite{dicke1954coherence}.
Using the fact that this sublattice rotation is a Gaussian unitary, we can analytically compute the MRE of the ground state of $H$ in this sector using the Dicke state, which provides a useful benchmark.
It behaves for large $L$ as~\cite{supplemental}
\begin{equation}\label{eq:dicke_asymptotic}
  M_2=\frac32\ln L-\ln\frac{4\sqrt2}{\pi^{3/2}}+\frac{9}{16L}+O(L^{-2}).
\end{equation}
We construct the exact MPS of the ground state at $\Delta=-1$ with bond dimension $D=L$.
In Fig.~\ref{fig:xxz_mre}(a), the MRE sampled from this MPS is in perfect agreement with Eq.~\eqref{eq:dicke_asymptotic} at all sizes up to $L=128$~\cite{supplemental}.

When $\Delta<-1$, our sampling method reveals a saturation of the MRE beyond the sizes accessible to exact diagonalization.
We obtain the ground state by density-matrix renormalization group (DMRG)~\cite{white1992density,white1993densitymatrix,schollwock2011densitymatrix}, project it onto the appropriate spin-flip eigenspace, and convert it into an MPS over the $2L$ modes~\cite{supplemental}.
As shown in Fig.~\ref{fig:xxz_mre}(a), the MRE saturates between $L=32$ and $64$ at $\Delta=-1.02$ and between $L=64$ and $128$ at $\Delta=-1.005$.
This saturation coincides with the emergence of a domain-wall profile in the spin-spin correlation function $\ev{S_1^zS_j^z}$; see Fig.~\ref{fig:xxz_mre}(b).
Since this correlator is a fermionic four-point function, the MRE resolves a change in many-body correlations that is inaccessible to the covariance matrix.

To interpret the saturation, we use the exact domain-wall ground states of the XXZ chain with suitable boundary fields~\cite{alcaraz1995anisotropic,koma1998complete}.
These fields select one of the two domain-wall orientations.
When the wall is well separated from the boundaries, the end spins in each orientation are nearly fully polarized, and the corresponding boundary fields act as a constant energy shift.
This motivates using an equal-weight superposition of the two orientations, in the appropriate spin-flip sector, as a reference state for the chain without boundary fields.
With $\cosh\eta=|\Delta|$, the MRE of this superposition has the small-$\eta$ expansion~\cite{supplemental}
\begin{equation}\label{eq:domain_wall_mre}
  M_2=\frac32\ln(\pi/\eta)+\ln4-\eta/2+O(\eta^2).
\end{equation}
This $L$-independent prediction agrees with the observed plateaus, linking MRE saturation to domain-wall formation.

The cost of these calculations is notably modest (see the Supplemental Material~\cite{supplemental}); at $\Delta=-1$, we confirmed that compressing the state at $L=64$ to $D=30$ leaves $M_2$ unchanged within the statistical error.
When $\Delta<-1$, the required bond dimension $D$ stops growing once the MRE saturates.
We also find that the required number of samples for fixed error scales as $\mathcal N\simeq e^{M_2}/\epsilon_M^2$. Consequently, it grows as $L^{3/2}$ at $\Delta=-1$, and the calculation for $L=128$ requires only about $1.4\times10^7$ samples for $\epsilon_M=0.01$.
When $\Delta<-1$, the required number of samples is an even smaller value owing to the saturation of the MRE (see the Supplemental Material~\cite{supplemental}).

%%%========== Conclusion ==========%%%
\textit{Conclusion.}---
We have established a practical method for efficiently computing the fermionic non-Gaussianity of many-body states represented as MPSs.
Our method enables a systematic study of many-body fermionic non-Gaussianity both in and out of equilibrium. For instance, the method can be used to determine the universal contribution to the MRE in critical ground states, which is fixed by the infrared boundary condition of the replicated conformal field theory~\cite{matsuda2026quantum}. For time-evolving states represented by MPSs, the same sampling procedure can track how non-Gaussian resources evolve after a quantum quench. Since the MRE is experimentally accessible via multiple copies of a state, numerical calculations can be directly compared with experimental data from fermionic quantum simulators, providing a benchmark for non-Gaussian resources realized in these platforms.

%%%=========== main text end ===========%%%

%%% acknowledgments %%%
\begin{acknowledgments}
  \textit{Acknowledgments.}---
  R.M. and M.H. were supported by FoPM, WINGS Program, the University of Tokyo.
  The work of M.H. was also supported by JSPS KAKENHI Grant No.~26KJ0910.
  Y.A. acknowledges support from JST FOREST Program (Grant No. JPMJFR222U), JST CREST (Grant No. JPMJCR23I2), and JST [Moonshot R\&D] (Grant No. JPMJMS256J). 
  The authors used ChatGPT (OpenAI), Codex (GPT-5.6-Sol and GPT-6-Astra, OpenAI), and Claude Code (Claude Opus 4.8, Opus 5, Opus 5.5, Fable 5, and Fable 5.1, Anthropic) for research and manuscript preparation. The authors checked all derivations, validated the sampling code against exact results, and revised AI-generated text. The authors take full responsibility for the paper.
\end{acknowledgments}

%%% bibliography %%%
\bibliography{biblio}

%%%=========== End Matter ===========%%%
\onecolumngrid
\begin{center}
  \textbf{\Large End Matter}
\end{center}
\beginendmatt

\textit{Derivation of Eq.~\eqref{eq:conditional_prob}.}---
We derive Eq.~\eqref{eq:conditional_prob} by explicitly summing over the output occupations at the sites that have not yet been sampled.
Define the product $A_{>j}^{\bm x}=A_{j+1}^{x_{j+1}}A_{j+2}^{x_{j+2}}\cdots A_n^{x_n}$.
For fixed input occupations $\bm x_{>j}$ and $\bm y_{>j}$ on sites $j+1,j+2,\ldots,n$, we define the partial contraction in which the prefix of copy $B$ has parity $p$ by
\begin{equation}
  G_p(\bm x_{>j},\bm y_{>j})
  =
  \left(A_{>j}^{\bm x}\right)^{\mathsf T}
  \mathcal B_j^{(p)}(\bm z_{\leq j})
  A_{>j}^{\bm y}.
\end{equation}
Splitting the amplitude at the bond after site $j$ gives
\begin{equation}\label{eq:end_suffix_split}
  \begin{aligned}
    \Phi(\bm z)
     & =
    s(\bm a,\bm b)
    \sum_{p=0}^1
    \sum_{\bm x_{>j},\bm y_{>j}}
    \\
     & \quad\times
    \left[
      \prod_{k=j+1}^n
      u_{z_k,2x_k+y_k}(-1)^{x_kp_{k-1}}
      \right]
    G_p(\bm x_{>j},\bm y_{>j}).
  \end{aligned}
\end{equation}
The local beam-splitter matrix obeys the following unitarity relation:
\begin{equation}
  \sum_{z_k=0}^3
  u_{z_k,2x_k+y_k}
  \overline{u_{z_k,2x_k'+y_k'}}
  =
  \delta_{x_kx_k'}\delta_{y_ky_k'}.
\end{equation}
Substituting Eq.~\eqref{eq:end_suffix_split} into the definition of $W_j$ and applying this relation gives the following expression:
\begin{equation}\label{eq:end_parity_cross_terms}
  \begin{aligned}
    W_j(\bm z_{\leq j})
     & =
    \sum_{p,p'=0}^1
    \sum_{\bm x_{>j},\bm y_{>j}}
    (-1)^{(p+p')\sum_{k>j}x_k}
    \\
     & \quad\times
    G_p(\bm x_{>j},\bm y_{>j})
    \overline{G_{p'}(\bm x_{>j},\bm y_{>j})}.
  \end{aligned}
\end{equation}
Since copy $B$ has fixed total parity $P$, $G_p(\bm x_{>j},\bm y_{>j})$ can be nonzero only if
\begin{equation}\label{eq:end_suffix_parity}
  \left(
  p+\sum_{k>j}y_k
  \right)\bmod2
  =
  P.
\end{equation}
This condition uniquely determines $p$ for each suffix configuration $\bm y_{>j}$, so the product of $G_p$ and $\overline{G_{p'}}$ vanishes whenever $p\neq p'$.
For the remaining diagonal terms with $p=p'$, the sign in Eq.~\eqref{eq:end_parity_cross_terms} equals unity, and repeated application of the right-canonical condition gives
\begin{equation}
  \sum_{\bm x_{>j}}
  A_{>j}^{\bm x}
  \left(A_{>j}^{\bm x}\right)^\dagger
  =
  I_{D_j}.
\end{equation}
Using this identity to evaluate the sums over $\bm x_{>j}$ and $\bm y_{>j}$ separately yields
\begin{equation}
  \begin{aligned}
    \sum_{\bm x_{>j},\bm y_{>j}}
    \left|G_p(\bm x_{>j},\bm y_{>j})\right|^2
     & =
    \Tr\!\left[
           \mathcal B_j^{(p)}(\bm z_{\leq j})
           \mathcal B_j^{(p)}(\bm z_{\leq j})^\dagger
           \right]
    \\
     & =
    \left\|\mathcal B_j^{(p)}(\bm z_{\leq j})\right\|_{\mathrm F}^2.
  \end{aligned}
\end{equation}
Finally, summing over the two diagonal parity sectors proves Eq.~\eqref{eq:conditional_prob}.

\textit{$U(1)$-resolved boundary recursion.}---
Particle-number conservation enables a more efficient sampling sweep.
We consider an input state satisfying $\psi(\bm x)=0$ unless $\sum_{j=1}^nx_j=N$ and assume that its MPS explicitly resolves the corresponding $U(1)$ symmetry.
The total input parity is $P=N\bmod2$.
To make the charge structure explicit, we decompose the virtual bond space after site $j$ as
\begin{align}
  \mathbb C^{D_j}
   & =
  \bigoplus_q\mathcal V_{j,q},
  \nonumber \\
  d_j(q)
   & =
  \dim\mathcal V_{j,q},
\end{align}
where $q$ counts the particles in the first $j$ sites.
With the boundary sectors $\mathcal V_{0,0}=\mathbb C$ and $\mathcal V_{n,N}=\mathbb C$, the $U(1)$ selection rule restricts each local tensor to blocks that connect left charge $q$ to right charge $q+x$:
\begin{equation}
  A_j^x(q)
  \in
  \mathbb C^{d_{j-1}(q)\times d_j(q+x)}.
\end{equation}

For a given prefix $\bm z_{\leq j}$, local particle-number conservation restricts the allowed charge blocks of the boundary matrix.
We denote the number of particles in the prefix by
\begin{align}
  K_j
   & =
  \sum_{i=1}^j(a_i+b_i),
  \nonumber \\
  K_0
   & =0,
\end{align}
and write $\mathcal B_j^{[q_1,q_2]}(\bm z_{\leq j})$ for the block of $\mathcal B_j^{(q_2\bmod2)}(\bm z_{\leq j})$ with row charge $q_1$ and column charge $q_2$.
Only blocks satisfying $q_1+q_2=K_j$ can be nonzero.
Starting from $\mathcal B_0^{[0,0]}=(1)$, Eq.~\eqref{eq:boundary_recursion} reduces to the block recursion
\begin{equation}\label{eq:u1_boundary_recursion}
  \begin{aligned}
    \mathcal B_j^{[Q_1,Q_2]}(\bm z_{\leq j})
     & =
    \sum_{\substack{
        q_1+q_2=K_{j-1}\\
        x,y\in\{0,1\}\\
        Q_1=q_1+x,\ Q_2=q_2+y
      }}
    u_{z_j,2x+y}
      (-1)^{x(q_2\bmod2)}
    \\
     & \quad\times
    \left[A_j^x(q_1)\right]^{\mathsf T}
    \mathcal B_{j-1}^{[q_1,q_2]}(\bm z_{<j})
    A_j^y(q_2).
  \end{aligned}
\end{equation}
Here, $q_2\bmod2$ is precisely the prefix parity $p_{j-1}$ of copy $B$, so no separate parity label is needed.
The beam splitter conserves the particle number at each site, and every nonzero matrix element of $u$ therefore satisfies $x+y=a_j+b_j$.
Together with $q_1+q_2=K_{j-1}$, this relation ensures that every updated block obeys $Q_1+Q_2=K_j$.

The perfect-sampling sweep retains the same probabilistic structure when expressed in terms of these symmetry-allowed charge blocks.
Because distinct charge blocks occupy orthogonal subspaces, Eq.~\eqref{eq:conditional_prob} becomes
\begin{equation}\label{eq:u1_prefix_weight}
  W_j(\bm z_{\leq j})
  =
  \sum_{q_1+q_2=K_j}
  \left\|
  \mathcal B_j^{[q_1,q_2]}(\bm z_{\leq j})
  \right\|_{\mathrm F}^2.
\end{equation}
For a trial value $z=2a+b$, the candidate boundary lies in the sector $K_j(z)=K_{j-1}+a+b$.
After $z_j$ is sampled, the charge blocks for the resulting prefix $\bm z_{\leq j}$ are normalized by
\begin{equation}
  \widehat{\mathcal B}_j^{[q_1,q_2]}(\bm z_{\leq j})
  =
  \frac{\mathcal B_j^{[q_1,q_2]}(\bm z_{\leq j})}
  {\sqrt{W_j(\bm z_{\leq j})}}.
\end{equation}
The update of $\lambda_j$ and the deterministic amplitude sweep then proceed exactly as in the dense formulation.
At the right boundary, the fixed charge of each copy leaves only the scalar block with $q_1=q_2=N$, and Eq.~\eqref{eq:normalized_amplitude} reduces to
\begin{equation}
  \Phi(\bm a,\bm b)
  =
  s(\bm a,\bm b)e^{\lambda_n}
  \left[\widehat{\mathcal B}_n^{[N,N]}(\bm z)\right]_{1,1}.
\end{equation}

Restricting the sweep to symmetry-allowed blocks can substantially reduce both the runtime and the working memory in practice.
For a prefix with charge $K_j$, the number of boundary elements that must be stored is
\begin{equation}
  \mathcal D_j(K_j)
  =
  \sum_{q_1+q_2=K_j}
  d_j(q_1)d_j(q_2),
\end{equation}
rather than the $2D_j^2$ elements of two dense parity boundaries.
The matrix multiplications in Eq.~\eqref{eq:u1_boundary_recursion} involve only charge-compatible blocks and can therefore be less expensive than the corresponding dense matrix multiplications.
The worst-case asymptotic scaling in terms of the total bond dimension remains $\mathcal O(nD^3)$ in time and $\mathcal O(D^2)$ in working memory per sweep, but the smaller charge blocks can make the method substantially more efficient in practice.
Particle-number conservation provides a further computational saving by allowing the sweeps for $\Phi(\bm a',\bm b)$ and $\Phi(\bm a,\bm b')$ to be skipped in some samples.
Each sample contains $2N$ particles in total, so if the two samples contain different numbers of particles in output $A$, these crossed amplitudes vanish, yielding $h(X)=0$ without either sweep.

%%%=========== Supplemental Material ===========%%%
\onecolumngrid
\clearpage
\begin{center}
  \textbf{\large Supplemental Material: Efficient Sampling for Many-Body Fermionic Non-Gaussianity}
\end{center}
\beginsupplement
\setlength{\tabcolsep}{10pt}

\section{\label{sec:sm_ground_state}Ground-state properties of the spinful XXZ chain}

This section collects the properties of the ground state of Eq.~\eqref{eq:xxz_chain} at $N_\uparrow=L/2$ that hold for every $\Delta$, namely the uniqueness, the spin-flip symmetry, and the covariance matrix.
As in the main text, all states in this section lie in the subspace with one fermion per site.

We first show that the ground state of $H$ at fixed $N_\uparrow$ is unique for every $\Delta$, as stated in the main text.
We use the sublattice rotation $U=\exp(i\pi\sum_{j:\mathrm{even}}n_{j\uparrow})$, which rotates the spins on the even sites by $\pi$ about the $z$ axis up to a constant phase.
It maps $c_{j\uparrow}$ to $-c_{j\uparrow}$ on the even sites and is therefore a number-conserving Gaussian unitary.
Since $US_j^\pm U^\dagger=-S_j^\pm$ on the even sites and every bond contains one even site, $U$ reverses the sign of $S_j^xS_{j+1}^x+S_j^yS_{j+1}^y$,
\begin{equation}
  UHU^\dagger=\sum_{j=1}^{L-1}\left(-S_j^xS_{j+1}^x-S_j^yS_{j+1}^y+\Delta S_j^zS_{j+1}^z\right).
  \label{eq:sm_rotated_hamiltonian}
\end{equation}
The off-diagonal elements of $UHU^\dagger$ in the $S^z$ basis are thus nonpositive, while $\Delta$ enters only the diagonal elements.
Each off-diagonal element exchanges antiparallel spins on neighboring sites, and a sequence of such exchanges connects every pair of configurations with the same $N_\uparrow$.
The Perron--Frobenius theorem then makes the ground state of $UHU^\dagger$ at fixed $N_\uparrow$ unique with positive amplitudes in the $S^z$ basis.
Since $U$ is unitary, the ground state of $H$ at fixed $N_\uparrow$ is unique as well.

Using this uniqueness, we next show that the ground state $\ket\psi$ of $H$ at $N_\uparrow=L/2$ is an eigenstate of the spin flip $\mathcal R=\prod_{j=1}^{L}(c_{j\uparrow}^\dagger c_{j\downarrow}+c_{j\downarrow}^\dagger c_{j\uparrow})$ with eigenvalue $(-1)^{L/2}$.
We choose the phase of $\ket\psi$ so that $U\ket\psi$ has positive amplitudes.
The operator $\mathcal R$ maps $(S_j^x,S_j^y,S_j^z)$ to $(S_j^x,-S_j^y,-S_j^z)$ and thus commutes with both $H$ and $UHU^\dagger$.
It maps the sector $N_\uparrow=L/2$ onto itself and permutes the $S^z$ basis states without signs.
The state $\mathcal RU\ket\psi$ is therefore a ground state of $UHU^\dagger$ with positive amplitudes, and the uniqueness gives $\mathcal RU\ket\psi=U\ket\psi$.
On the states with one fermion per site, $n_{j\downarrow}=1-n_{j\uparrow}$ gives $\mathcal RU\mathcal R=(-1)^{L/2}U$, since there are $L/2$ even sites.
Combining these relations with $U^2=I$, we obtain
\begin{equation}
  \mathcal R\ket\psi=\mathcal RU\,U\ket\psi=(-1)^{L/2}U\mathcal R\,U\ket\psi=(-1)^{L/2}\ket\psi.
  \label{eq:sm_spin_flip_eigenvalue}
\end{equation}

Using this spin-flip symmetry, we finally derive the value of $\ev{c_{i\sigma}^\dagger c_{j\sigma'}}$ stated in the main text and show that the Majorana covariance matrix of $\ket\psi$ vanishes for every $\Delta$.
We write $C_{ij}=\ev{c_i^\dagger c_j}$ and $\kappa_{ij}=\ev{c_ic_j}$ for the correlation matrices in the mode basis and $\Gamma_{\mu\nu}=\frac{i}{2}\ev{[\gamma_\mu,\gamma_\nu]}$ for the Majorana covariance matrix.
The fermionic antiflatness of Ref.~\cite{sierant2026fermionic} is $\mathcal F_k=n-\frac12\Tr[(\Gamma^{\mathsf T}\Gamma)^k]$ for $n$ modes, which vanishes for Gaussian states and takes its maximal value $n$ when $\Gamma=0$.
The argument below uses only the fact that the state has one fermion per site with $N_\uparrow=L/2$ and is an eigenstate of $\mathcal R$.
The correlation $\ev{c_{i\sigma}^\dagger c_{j\sigma'}}$ with $i\neq j$ is the expectation value of an operator that changes the particle numbers of two sites and therefore vanishes.
The on-site correlation $\ev{c_{j\uparrow}^\dagger c_{j\downarrow}}=\ev{S_j^+}$ changes $N_\uparrow$ and vanishes as well.
The spin-flip symmetry gives $\ev{n_{j\uparrow}}=\ev{n_{j\downarrow}}=1/2$, and particle-number conservation gives $\kappa=0$.
Hence
\begin{equation}
  C=\frac{1}{2}I_{2L},\qquad \kappa=0,\qquad \Gamma=0,\qquad \mathcal F_k=2L,
  \label{eq:sm_covariance}
\end{equation}
independently of $\Delta$ and $L$.

\section{\label{sec:sm_analytic}Analytical results for the MRE in the XXZ chain}

This section derives Eqs.~\eqref{eq:dicke_asymptotic} and~\eqref{eq:domain_wall_mre} of the main text.
Section~\ref{subsec:sm_dicke_exact} obtains the exact MRE of the ground state at $\Delta=-1$ for every $L$.
Section~\ref{subsec:sm_asymptotic} derives the asymptotic forms at $\Delta=-1$ and $\Delta<-1$.
Since $U$ is a Gaussian unitary, we perform all calculations with the rotated Hamiltonian $UHU^\dagger$ of Eq.~\eqref{eq:sm_rotated_hamiltonian}.

\subsection{\label{subsec:sm_dicke_exact}Exact MRE at \texorpdfstring{$\Delta=-1$}{Delta=-1}}

At $\Delta=-1$, the rotated Hamiltonian $UHU^\dagger=-\sum_j\bm S_j\cdot\bm S_{j+1}$ is the ferromagnetic Heisenberg chain, and the state with all spins up attains the lowest eigenvalue $-1/4$ of every term.
The total lowering operator commutes with this Hamiltonian, so the Dicke state~\cite{dicke1954coherence}
\begin{equation}
  \ket{\mathrm D_L}=\binom{L}{L/2}^{-1/2}\sum_{\abs{\bm s}=L/2}\ket{\bm s},
  \label{eq:sm_dicke_state}
\end{equation}
obtained by applying it $L/2$ times, is also a ground state.
Here $\ket{\bm s}$ is the state with one fermion per site in the spin configuration $\bm s$, and $\abs{\bm s}$ is its number of up spins.
By the uniqueness shown in Sec.~\ref{sec:sm_ground_state}, $\ket{\mathrm D_L}$ is the unique ground state at $N_\uparrow=L/2$.

We regard the amplitudes $\Phi(\bm a,\bm b)$ of the main text as a matrix $\Phi$ with rows $\bm a$ and columns $\bm b$.
Since $\mathcal C(\rho)=\Phi\Phi^\dagger$, the purity $\nu_2=\Tr[(\Phi\Phi^\dagger)^2]$ is the sum of the fourth powers of the singular values of $\Phi$.
As shown below, $\Phi$ splits into blocks labeled by the numbers of fermions at each site and a magnetization, and the singular values of each block depend only on the magnitude and the $z$ component of the total spin.

The beam splitter $U_{\rm BS}$ mixes each mode only with the same mode of the other copy and therefore acts site by site.
We label the output at site $j$ by the pair $(a_j,b_j)$ of the states of copies $A$ and $B$.
Here $a_j,b_j\in\{0,\uparrow,\downarrow,\uparrow\downarrow\}$, where $0$ and $\uparrow\downarrow$ denote an empty and a doubly occupied site.
We write $\ket{a,b}$ for the state of one site with the creation operators ordered as $c_{A\uparrow}^\dagger,c_{A\downarrow}^\dagger,c_{B\uparrow}^\dagger,c_{B\downarrow}^\dagger$.
From $U_{\rm BS}c_{A\sigma}^\dagger U_{\rm BS}^\dagger=(c_{A\sigma}^\dagger-c_{B\sigma}^\dagger)/\sqrt2$ and $U_{\rm BS}c_{B\sigma}^\dagger U_{\rm BS}^\dagger=(c_{A\sigma}^\dagger+c_{B\sigma}^\dagger)/\sqrt2$, we obtain
\begin{equation}
  \begin{aligned}
    U_{\rm BS}\ket{\uparrow,\uparrow}&=\ket{\uparrow,\uparrow},\qquad
    U_{\rm BS}\ket{\downarrow,\downarrow}=\ket{\downarrow,\downarrow},\\
    U_{\rm BS}\ket{\uparrow,\downarrow}&=\tfrac12\left(\ket{\uparrow\downarrow,0}+\ket{\uparrow,\downarrow}+\ket{\downarrow,\uparrow}-\ket{0,\uparrow\downarrow}\right),\\
    U_{\rm BS}\ket{\downarrow,\uparrow}&=\tfrac12\left(-\ket{\uparrow\downarrow,0}+\ket{\uparrow,\downarrow}+\ket{\downarrow,\uparrow}+\ket{0,\uparrow\downarrow}\right).
  \end{aligned}
  \label{eq:sm_site_bs}
\end{equation}
To apply this relation at every site, we reorder the creation operators, which are ordered copy by copy, so that the four modes of each site are adjacent.
The reorderings of the input and the output give the signs $(-1)^{L(L-1)/2}$ and $(-1)^{\sum_{j<j'}n(b_j)n(a_{j'})}$, where $n(a)$ is the number of fermions in the state $a$.
For an input state $\ket\psi$ with one fermion per site, we thus obtain
\begin{equation}
  \Phi(\bm a,\bm b)=(-1)^{L(L-1)/2}(-1)^{\sum_{j<j'}n(b_j)n(a_{j'})}\sum_{\bm s_1,\bm s_2}\psi(\bm s_1)\psi(\bm s_2)\prod_{j=1}^{L}\bra{a_j,b_j}U_{\rm BS}\ket{s_{1j},s_{2j}},
  \label{eq:sm_site_amplitude}
\end{equation}
where $\psi(\bm s)=\braket{\bm s}{\psi}$, $\bm s_1$ and $\bm s_2$ are the input spin configurations of copies $A$ and $B$, and $s_{ij}$ is the spin of $\bm s_i$ at site $j$.
The signs do not depend on the spin orientations.
The beam splitter conserves the number of fermions at each site and the total number of up spins of the two copies.
For the input $\ket{\mathrm D_L}$, $\Phi(\bm a,\bm b)$ is therefore nonzero only if $n(a_j)+n(b_j)=2$ at every site and the output contains $L$ up spins in total.
Here a doubly occupied site counts as containing one up spin.

Let $k$ be the number of sites at which copy $A$ holds one fermion.
At each of the other $L-k$ sites, one copy is doubly occupied and the other is empty, so the site contains one up spin.
If copy $A$ holds $k/2+m$ up spins at the $k$ sites, copy $B$ therefore holds $k/2-m$ up spins there.
The rows and columns with fixed numbers of fermions at each site and fixed $m$ form a block of $\Phi$, and the reordering sign is constant within the block.
The rows of the block are the spin configurations of copy $A$ at the $k$ sites, and the columns are those of copy $B$.
If we label each column by the configuration with all spins of copy $B$ flipped, both rows and columns become configurations of $k$ spins with $k/2+m$ up spins (Fig.~\ref{fig:sm_block}).
We hereafter regard each block as a matrix on the state space of these $k$ spins $1/2$.

To find the entries of the blocks, we write $\ket{\mathrm D_L}$ as a projected product state.
The state $\ket{\mathrm D_L}$ is the projection onto $N_\uparrow=L/2$ of the product state with all spins along the $x$ direction, and writing the projection as an integral over a phase gives
\begin{equation}
  \ket{\mathrm D_L}=\binom{L}{L/2}^{-1/2}\int_0^{2\pi}\frac{d\theta}{2\pi}\,e^{-iL\theta/2}\bigotimes_{j=1}^{L}\left(\ket\downarrow+e^{i\theta}\ket\uparrow\right).
  \label{eq:sm_dicke_phase}
\end{equation}
Indeed, the coefficient of $\ket{\bm s}$ in the expanded product is $e^{i\theta\abs{\bm s}}$, and the integral over $\theta$ keeps only the terms with $\abs{\bm s}=L/2$.
For the two copies, we use this representation with the phases $\theta+\delta$ and $\theta-\delta$ and write
\begin{equation}
  \ket{\mathrm D_L}_A\ket{\mathrm D_L}_B=\binom{L}{L/2}^{-1}\int_0^{2\pi}\frac{d\theta}{2\pi}\int_0^{2\pi}\frac{d\delta}{2\pi}\,e^{-iL\theta}\bigotimes_{j=1}^{L}\left(\ket\downarrow+e^{i(\theta+\delta)}\ket\uparrow\right)_A\bigotimes_{j=1}^{L}\left(\ket\downarrow+e^{i(\theta-\delta)}\ket\uparrow\right)_B.
  \label{eq:sm_dicke_two_copy}
\end{equation}
The integral over $\theta$ restricts the sum of the numbers of up spins of the two copies to $L$, and the integral over $\delta$ restricts their difference to $0$, so this indeed gives $\ket{\mathrm D_L}_A\ket{\mathrm D_L}_B$.
When this representation is inserted into Eq.~\eqref{eq:sm_site_amplitude}, the sums over $\bm s_1$ and $\bm s_2$ factorize over sites, and we obtain
\begin{equation}
  \Phi(\bm a,\bm b)=\pm\binom{L}{L/2}^{-1}\int_0^{2\pi}\frac{d\theta}{2\pi}\int_0^{2\pi}\frac{d\delta}{2\pi}\,e^{-iL\theta}\prod_{j=1}^{L}f_{a_jb_j}(\theta,\delta).
  \label{eq:sm_dicke_amplitude}
\end{equation}
Here $\pm$ denotes the sign in Eq.~\eqref{eq:sm_site_amplitude}, and $f_{a_jb_j}(\theta,\delta)=\bra{a_j,b_j}U_{\rm BS}\big(\ket{\downarrow,\downarrow}+e^{i(\theta+\delta)}\ket{\uparrow,\downarrow}+e^{i(\theta-\delta)}\ket{\downarrow,\uparrow}+e^{2i\theta}\ket{\uparrow,\uparrow}\big)$ is the factor of site $j$.
Applying Eq.~\eqref{eq:sm_site_bs} to this input gives
\begin{equation}
  \renewcommand{\arraystretch}{1.3}
  \begin{array}{c|cccccc}
    (a_j,b_j)&(\uparrow,\uparrow)&(\downarrow,\downarrow)&(\uparrow,\downarrow)&(\downarrow,\uparrow)&(\uparrow\downarrow,0)&(0,\uparrow\downarrow)\\\hline
    f_{a_jb_j}(\theta,\delta)&e^{2i\theta}&1&e^{i\theta}\cos\delta&e^{i\theta}\cos\delta&ie^{i\theta}\sin\delta&-ie^{i\theta}\sin\delta
  \end{array}
  \label{eq:sm_dicke_site}
\end{equation}
For example, the factor of $(\uparrow,\downarrow)$ is the sum $[e^{i(\theta+\delta)}+e^{i(\theta-\delta)}]/2=e^{i\theta}\cos\delta$ of the contributions from the inputs $\ket{\uparrow,\downarrow}$ and $\ket{\downarrow,\uparrow}$.
Every factor contains $e^{i\theta}$ to the power of the number of up spins at the site.
Since an entry of a block has $L$ up spins in total, these powers cancel $e^{-iL\theta}$, and the integral over $\theta$ gives one.
What remains is the integral over $\delta$ of the $\delta$-dependent parts of the factors in Eq.~\eqref{eq:sm_dicke_site}.

Within a block, the product of the $\delta$-dependent parts of the $L-k$ remaining sites is $i^{L-k}\sin^{L-k}\delta$ up to a sign.
At each of the $k$ sites, the $\delta$-dependent part is $\cos\delta$ when the spins of the row and the column are equal, that is, when the output is $(\uparrow,\downarrow)$ or $(\downarrow,\uparrow)$, and $1$ when they differ.
These are the matrix elements of the spin-$1/2$ operator $\cos\delta+2S_j^x$.
Each block is therefore equal, up to a sign, to the operator
\begin{equation}
  \binom{L}{L/2}^{-1}i^{L-k}\int_0^{2\pi}\frac{d\delta}{2\pi}\,\sin^{L-k}\delta\prod_{j}\left(\cos\delta+2S_j^x\right)
  \label{eq:sm_dicke_block}
\end{equation}
projected onto the subspace in which the $z$ component of the total spin is $m$.
Here the product runs over the $k$ sites, and the total spin refers to these $k$ spins.
For odd $k$, replacing $\delta$ with $-\delta$ reverses the sign of the integrand, so the block vanishes.

\begin{figure}[htbp]
  \centering
  \begin{tikzpicture}[x=1.5cm,y=0.8cm,
      cell/.style={draw,minimum width=1.1cm,minimum height=0.6cm,inner sep=0pt,font=\small}]
    \foreach \j in {2,5}{\fill[gray!20] (\j-0.5,0.5) rectangle (\j+0.5,-3.5);}
    \foreach \j in {1,...,6}{\node[font=\small] at (\j,0.9) {$\j$};}
    \node[anchor=east,font=\small] at (0.4,0.9) {site};
    \node[anchor=east,font=\small] at (0.4,0) {copy $A$};
    \node[anchor=east,font=\small] at (0.4,-1) {copy $B$};
    \node[anchor=east,font=\small] at (0.4,-2) {$B$ flipped};
    \node[anchor=east,font=\small] at (0.4,-3) {$\delta$-dependent part};
    \foreach \j/\lab in {1/\uparrow,2/\uparrow\downarrow,3/\downarrow,4/\uparrow,5/0,6/\downarrow}{\node[cell] at (\j,0) {$\lab$};}
    \foreach \j/\lab in {1/\downarrow,2/0,3/\downarrow,4/\uparrow,5/\uparrow\downarrow,6/\uparrow}{\node[cell] at (\j,-1) {$\lab$};}
    \foreach \j/\lab in {1/\uparrow,3/\uparrow,4/\downarrow,6/\downarrow}{\node[cell] at (\j,-2) {$\lab$};}
    \foreach \j/\fac in {1/\cos\delta,2/i\sin\delta,3/1,4/1,5/-i\sin\delta,6/\cos\delta}{\node[font=\small] at (\j,-3) {$\fac$};}
  \end{tikzpicture}
  \caption{Example of an entry of $\Phi$ for $L=6$.
    At the gray sites, copy $A$ is doubly occupied or empty, and at the other $k=4$ sites each copy holds one fermion.
    The row configuration is $(\uparrow,\downarrow,\uparrow,\downarrow)$ of copy $A$, and the column configuration is $(\uparrow,\uparrow,\downarrow,\downarrow)$ of copy $B$ after flipping, both with two up spins ($m=0$).
    The bottom row shows the $\delta$-dependent parts of the factors in Eq.~\eqref{eq:sm_dicke_site}.
    At the $k$ sites, they are $\cos\delta$ when the spins of the row and the column are equal and $1$ when they differ.}
  \label{fig:sm_block}
\end{figure}

The product in Eq.~\eqref{eq:sm_dicke_block} is diagonal in the eigenstates of the individual $S_j^x$ and takes the value $(\cos\delta+1)^{k/2+m_x}(\cos\delta-1)^{k/2-m_x}$ on a state in which the $x$ component of the total spin is $m_x$.
The product is thus a function of the $x$ component of the total spin and preserves each multiplet of total spin $S$.
Each multiplet contains exactly one state $\ket{S,m}$ in which the $z$ component of the total spin is $m$, so the states $\ket{S,m}$ of all multiplets are eigenvectors of the projected block.
The diagonal element $\bra{S,m}\prod_j(\cos\delta+2S_j^x)\ket{S,m}$ that gives the eigenvalue is a matrix element of a function of the total-spin operators and is therefore the same for all multiplets.
We evaluate it in the multiplet in which $k-2S$ spins form $k/2-S$ singlet pairs and the other $2S$ spins are in a symmetric state.
On a singlet pair $(j,j')$, the operator $(\cos\delta+2S_j^x)(\cos\delta+2S_{j'}^x)$ has the eigenvalue $\cos^2\delta-1=-\sin^2\delta$.
The state of the $2S$ symmetric spins with the $z$ component $m$ is the Dicke state, that is, the equal-weight superposition of all configurations with $S+m$ up spins.
In the expansion of the product $\prod_j(\cos\delta+2S_j^x)$, each term flips some of the spins and multiplies the others by $\cos\delta$.
In the diagonal element for the Dicke state, the $\binom{S+m}{g}\binom{S-m}{g}$ terms that flip $g$ of the $S+m$ up spins and $g$ of the $S-m$ down spins of one configuration each contribute $\cos^{2S-2g}\delta$.
We thus obtain
\begin{equation}
  \bra{S,m}\prod_j\left(\cos\delta+2S_j^x\right)\ket{S,m}=\left(-\sin^2\delta\right)^{k/2-S}\sum_{g=0}^{S-\abs{m}}\binom{S+m}{g}\binom{S-m}{g}\cos^{2S-2g}\delta.
  \label{eq:sm_dicke_diag}
\end{equation}

The factors in Eqs.~\eqref{eq:sm_dicke_block} and~\eqref{eq:sm_dicke_diag} combine into $i^{L-k}\sin^{L-k}\delta\,(-\sin^2\delta)^{k/2-S}=(-1)^{L/2-S}\sin^{L-2S}\delta$, which does not depend on $k$.
Each block is real and symmetric, so its singular values are the absolute values of its eigenvalues.
Using the beta-function integral
\begin{equation}
  \int_0^{2\pi}\frac{d\delta}{2\pi}\,\sin^{L-2S}\delta\cos^{2S-2g}\delta=\frac1\pi B\Big(\frac{L-2S+1}{2},\frac{2S-2g+1}{2}\Big)=\frac{(L-2S)!\,(2S-2g)!}{4^{L/2-g}\,(L/2-S)!\,(S-g)!\,(L/2-g)!},
\end{equation}
we obtain the singular values
\begin{equation}
  \Lambda_{S,m}=\binom{L}{L/2}^{-1}\sum_{g=0}^{S-\abs{m}}\binom{S+m}{g}\binom{S-m}{g}\frac{(L-2S)!\,(2S-2g)!}{4^{L/2-g}\,(L/2-S)!\,(S-g)!\,(L/2-g)!}.
  \label{eq:sm_dicke_lambda}
\end{equation}

\begin{figure}[htbp]
  \centering
  \includegraphics{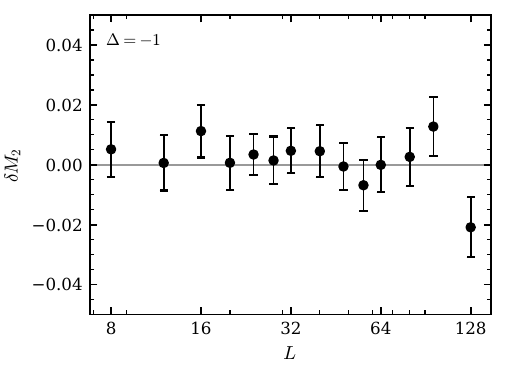}
  \caption{Deviation $\delta M_2=\widetilde M_2-M_2(\mathrm D_L)$ of the MRE $\widetilde M_2$ sampled from the exact MPS at $\Delta=-1$ in Fig.~\ref{fig:xxz_mre}(a) from the exact value of Eq.~\eqref{eq:sm_dicke_exact}.
    The error bars are one standard error.}
  \label{fig:sm_dicke_benchmark}
\end{figure}

Since $\Lambda_{S,m}$ depends only on $S$ and $m$, it remains to count the multiplets of total spin $S$ in all blocks.
Empty and doubly occupied sites have zero spin, so this number equals the number of multiplets of total spin $S$ of copy $A$ in the space spanned by all output configurations of copy $A$.
If we assign $1,1,z,z^{-1}$ to the site states $0,\uparrow\downarrow,\uparrow,\downarrow$, the number of configurations in which the $z$ component of the total spin is $m$ is the coefficient $\binom{2L}{L+2m}$ of $z^{2m}$ in $(2+z+z^{-1})^L=z^{-L}(1+z)^{2L}$.
The number of multiplets of total spin $S$ is the difference between the numbers of configurations with the $z$ components $S$ and $S+1$, which gives
\begin{equation}
  \nu_2(\mathrm D_L)=\sum_{S=0}^{L/2}\left[\binom{2L}{L+2S}-\binom{2L}{L+2S+2}\right]\sum_{m=-S}^{S}\Lambda_{S,m}^4.
  \label{eq:sm_dicke_exact}
\end{equation}
Half-integer values of $S$ arise only from odd $k$, so only integer values of $S$ appear in the sum.
Figure~\ref{fig:sm_dicke_benchmark} compares this exact value with the MRE sampled from the exact MPS in Fig.~\ref{fig:xxz_mre}(a), and the deviation lies within $2.1$ standard errors at all sizes.

\subsection{\label{subsec:sm_asymptotic}Asymptotic forms of the MRE}

\subsubsection{\label{subsec:sm_expansion}Asymptotic expansion of the purity of number-projected product states}

The state $\ket{\mathrm D_L}$ and the domain-wall state of Eq.~\eqref{eq:sm_kink_state} below are both of the form
\begin{equation}
  \ket\xi=\frac{1}{\sqrt Z}P_{N_\uparrow=L/2}\bigotimes_{j=1}^{L}\left(\sqrt{p_j}\ket\uparrow+\sqrt{1-p_j}\ket\downarrow\right),
  \label{eq:sm_projected_state}
\end{equation}
where $P_{N_\uparrow=L/2}$ is the projector onto $N_\uparrow=L/2$, $0<p_j<1$, and $Z=\sum_{\abs{\bm s}=L/2}\Pr(\bm s)$ is the probability of $N_\uparrow=L/2$ in the product state.
Here $\Pr(\bm s)=\prod_{j=1}^{L}p_j^{s_j}(1-p_j)^{1-s_j}$ is the probability of the spin configuration $\bm s$ in the product state, where $s_j$ is $1$ for up and $0$ for down.
In this section, we derive an asymptotic expansion of the purity $\nu_2(\xi)$ of $\ket\xi$ up to first order in the inverse of the variance of $N_\uparrow$ in the product state.

The purity of a state of the form of Eq.~\eqref{eq:sm_projected_state} can be written as a probability.
We first insert Eq.~\eqref{eq:sm_site_amplitude} into Eq.~\eqref{nu2} of the main text for a state $\ket\psi$ with one fermion per site.
In a nonzero term, $n(a'_j)=2-n(b_j)=n(a_j)$ and $n(b'_j)=n(b_j)$, so the four amplitudes carry the same sign, whose fourth power is one.
The sum over the output configurations factorizes over sites, which gives
\begin{equation}
  \nu_2(\psi)=\sum_{\bm s_1,\ldots,\bm s_4}\;\sum_{\bm t_1,\ldots,\bm t_4}\;\prod_{i=1}^{4}\overline{\psi(\bm t_i)}\,\psi(\bm s_i)\prod_{j=1}^{L}V_{t^{(j)}s^{(j)}}.
  \label{eq:sm_nu2_V}
\end{equation}
Here $(\bm t_1,\bm t_2)$ and $(\bm t_3,\bm t_4)$ are the input spin configurations of the complex-conjugated amplitudes $\Phi(\bm a,\bm b)$ and $\Phi(\bm a',\bm b')$, and $(\bm s_1,\bm s_2)$ and $(\bm s_3,\bm s_4)$ are those of $\Phi(\bm a',\bm b)$ and $\Phi(\bm a,\bm b')$.
The coefficients in Eq.~\eqref{eq:sm_site_bs} are real, so the complex conjugation acts only on $\psi(\bm t_i)$.
The four spins of these configurations at site $j$ are $s^{(j)}=(s_{1j},\ldots,s_{4j})$ and $t^{(j)}=(t_{1j},\ldots,t_{4j})$, and the $16\times16$ matrix on the four spins $s$ and $t$ of one site is
\begin{equation}
  V_{ts}=\sum_{a,b,a',b'}\bra{a,b}U_{\rm BS}\ket{t_1,t_2}\bra{a',b'}U_{\rm BS}\ket{t_3,t_4}\bra{a',b}U_{\rm BS}\ket{s_1,s_2}\bra{a,b'}U_{\rm BS}\ket{s_3,s_4}.
  \label{eq:sm_v_matrix}
\end{equation}

Let $\abs{s}$ be the number of up spins in $s$.
By using Eq.~\eqref{eq:sm_site_bs}, the nonzero entries of $V$ are obtained as follows.
For $s={\uparrow\uparrow\uparrow\uparrow}$ and $s={\downarrow\downarrow\downarrow\downarrow}$, the only nonzero entry in the column of $s$ is $V_{ss}=1$.
For $\abs{s}=1$ and $\abs{s}=3$, $V_{ts}=1/4$ for each of the four states $t$ with $\abs{t}=\abs{s}$.
For $\abs{s}=2$, the entries in the basis ${\downarrow\downarrow\uparrow\uparrow}$, ${\downarrow\uparrow\downarrow\uparrow}$, ${\downarrow\uparrow\uparrow\downarrow}$, ${\uparrow\downarrow\downarrow\uparrow}$, ${\uparrow\downarrow\uparrow\downarrow}$, ${\uparrow\uparrow\downarrow\downarrow}$ are
\begin{equation}
  \frac14\begin{pmatrix}
    0&1&1&1&1&0\\
    1&1&0&0&1&1\\
    1&0&1&1&0&1\\
    1&0&1&1&0&1\\
    1&1&0&0&1&1\\
    0&1&1&1&1&0
  \end{pmatrix}.
  \label{eq:sm_v_block}
\end{equation}
These entries show that $V$ satisfies
\begin{gather}
  V_{ts}\geq0,\qquad\sum_tV_{ts}=1,\label{eq:sm_v_column}\\
  \sum_sV_{ts}=1,\label{eq:sm_v_row}\\
  V_{ts}=0\quad(\abs{t}\neq\abs{s}),\label{eq:sm_v_conserve}\\
  V_{\bar t\bar s}=V_{ts},\label{eq:sm_v_flip}\\
  \sum_{t:\,t_i=\uparrow}V_{ts}=\frac{\abs{s}}{4}\quad(i=1,\ldots,4),\label{eq:sm_v_marginal}
\end{gather}
where $\bar s$ is the configuration obtained from $s$ by flipping all spins.

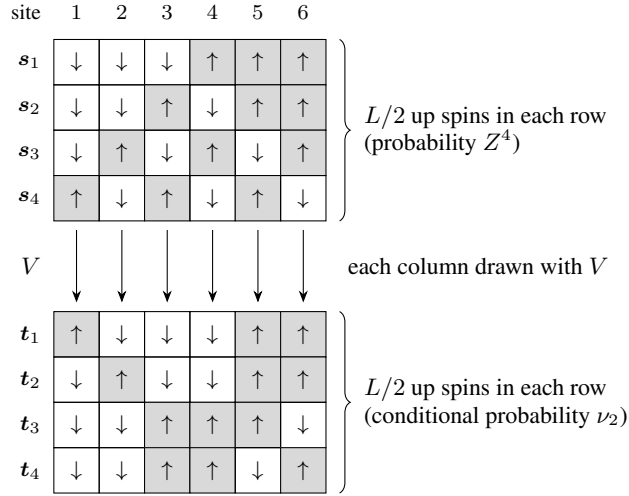
\begin{figure}[htbp]
  \centering
  \begin{tikzpicture}[x=6mm,y=6mm]
    \foreach \row [count=\i] in {{0,0,0,1,1,1},{0,0,1,0,1,1},{0,1,0,1,0,1},{1,0,1,0,1,0}}{
      \foreach \bitval [count=\j] in \row{
        \ifnum\bitval=1
          \fill[gray!30] (\j-1,-\i) rectangle (\j,1-\i);
          \node[font=\footnotesize] at (\j-0.5,0.5-\i) {$\uparrow$};
        \else
          \node[font=\footnotesize] at (\j-0.5,0.5-\i) {$\downarrow$};
        \fi
        \draw (\j-1,-\i) rectangle (\j,1-\i);
      }
      \node[anchor=east,font=\footnotesize] at (-0.1,0.5-\i) {$\bm s_{\i}$};
    }
    \foreach \row [count=\i] in {{1,0,0,0,1,1},{0,1,0,0,1,1},{0,0,1,1,1,0},{0,0,1,1,0,1}}{
      \foreach \bitval [count=\j] in \row{
        \ifnum\bitval=1
          \fill[gray!30] (\j-1,-6-\i) rectangle (\j,-5-\i);
          \node[font=\footnotesize] at (\j-0.5,-5.5-\i) {$\uparrow$};
        \else
          \node[font=\footnotesize] at (\j-0.5,-5.5-\i) {$\downarrow$};
        \fi
        \draw (\j-1,-6-\i) rectangle (\j,-5-\i);
      }
      \node[anchor=east,font=\footnotesize] at (-0.1,-5.5-\i) {$\bm t_{\i}$};
    }
    \foreach \j in {1,...,6}{
      \node[font=\footnotesize] at (\j-0.5,0.6) {$\j$};
      \draw[-{Stealth[length=1.6mm]}] (\j-0.5,-4.2) -- (\j-0.5,-5.8);
    }
    \node[anchor=east,font=\footnotesize] at (-0.1,0.6) {site};
    \node[anchor=east,font=\small] at (-0.1,-5) {$V$};
    \draw[decorate,decoration={brace,amplitude=4pt}] (6.3,0) -- (6.3,-4)
      node[midway,right=6pt,align=left,font=\small]{$L/2$ up spins in each row\\(probability $Z^4$)};
    \node[anchor=west,align=left,font=\small] at (6.3,-5) {each column drawn with $V$};
    \draw[decorate,decoration={brace,amplitude=4pt}] (6.3,-6) -- (6.3,-10)
      node[midway,right=6pt,align=left,font=\small]{$L/2$ up spins in each row\\(conditional probability $\nu_2$)};
  \end{tikzpicture}
  \caption{Example of the event in Eq.~\eqref{eq:sm_probability} for $L=6$.
    The upper four rows are $\bm s_1,\ldots,\bm s_4$, the lower four rows are $\bm t_1,\ldots,\bm t_4$, and column $j$ shows the four spins $s^{(j)}$ and $t^{(j)}$ at site $j$.
    The spins in the upper rows are drawn independently, and each lower column is drawn from the upper column with probability $V_{t^{(j)}s^{(j)}}$.
    In this example, every column has $V_{t^{(j)}s^{(j)}}=1/4$, and all eight rows have three up spins.}
  \label{fig:sm_arrays}
\end{figure}

The amplitudes of the state in Eq.~\eqref{eq:sm_projected_state} are $\xi(\bm s)=Z^{-1/2}\delta_{\abs{\bm s},L/2}\sqrt{\Pr(\bm s)}$.
The product $\prod_i\Pr(\bm s_i)=\prod_jp_j^{\abs{s^{(j)}}}(1-p_j)^{4-\abs{s^{(j)}}}$ depends only on the numbers $\abs{s^{(j)}}$ of up spins at each site.
In the nonzero terms of Eq.~\eqref{eq:sm_nu2_V}, Eq.~\eqref{eq:sm_v_conserve} gives $\abs{t^{(j)}}=\abs{s^{(j)}}$, so $\prod_i\Pr(\bm t_i)=\prod_i\Pr(\bm s_i)$ and the square roots disappear.
Furthermore, by Eq.~\eqref{eq:sm_v_column}, $\prod_i\Pr(\bm s_i)\prod_jV_{t^{(j)}s^{(j)}}$ is the probability of a process in which $\bm s_1,\ldots,\bm s_4$ are drawn independently from the product state and $t^{(j)}$ is then drawn at each site with probability $V_{t^{(j)}s^{(j)}}$ (Fig.~\ref{fig:sm_arrays}).
The probability that each of $\bm s_1,\ldots,\bm s_4$ has $L/2$ up spins is $Z^4$, so
\begin{equation}
  \nu_2(\xi)=\Pr\left[\abs{\bm t_i}=L/2\ (i=1,\ldots,4)\,\middle|\,\abs{\bm s_i}=L/2\ (i=1,\ldots,4)\right].
  \label{eq:sm_probability}
\end{equation}

We expand Eq.~\eqref{eq:sm_probability} when the variance
\begin{equation}
  v=\sum_{j=1}^{L}p_j(1-p_j)
  \label{eq:sm_variance}
\end{equation}
of $N_\uparrow$ in the product state is large.
We assume $p_{L+1-j}=1-p_j$, as is the case for $\ket{\mathrm D_L}$ and for the domain-wall state of Eq.~\eqref{eq:sm_kink_state}.
This assumption implies $\sum_jp_j=L/2$.
We collect the seven spins $s_{1j},s_{2j},s_{3j},s_{4j},t_{1j},t_{2j},t_{3j}$ at site $j$ into a vector $Y_j\in\{0,1\}^7$, with $1$ for up and $0$ for down.
By Eq.~\eqref{eq:sm_v_conserve}, the condition on $\bm t_4$ follows from the other seven conditions.
Equation~\eqref{eq:sm_probability} therefore becomes
\begin{equation}
  \nu_2(\xi)=Z^{-4}\Pr\Big[\sum_jY_j=\tfrac L2(1,\ldots,1)\Big].
  \label{eq:sm_probability_Y}
\end{equation}
Note that the vectors $Y_j$ of different sites are independent.

At site $j$, the spins $s_{1j},\ldots,s_{4j}$ are independent, each with mean $p_j$ and variance $p_j(1-p_j)$.
Since the probability of $s^{(j)}=s$ is $p_j^{\abs{s}}(1-p_j)^{4-\abs{s}}$, the probability of $t^{(j)}=t$ is
\begin{equation}
  \sum_sV_{ts}\,p_j^{\abs{s}}(1-p_j)^{4-\abs{s}}=p_j^{\abs{t}}(1-p_j)^{4-\abs{t}}\sum_sV_{ts}=p_j^{\abs{t}}(1-p_j)^{4-\abs{t}}.
\end{equation}
Hence $t^{(j)}$ has the same distribution as $s^{(j)}$.
The first equality uses Eq.~\eqref{eq:sm_v_conserve}, and the second uses Eq.~\eqref{eq:sm_v_row}.
By Eq.~\eqref{eq:sm_v_marginal}, the spin $t_{ij}$ is up with probability $\abs{s^{(j)}}/4$ for given $s^{(j)}$, so the covariance between $s_{kj}$ and $t_{ij}$ is $p_j(1-p_j)/4$.
The covariance matrix of $Y_j$ is thus $p_j(1-p_j)\Sigma_0$ with
\begin{equation}
  \Sigma_0=\begin{pmatrix}I_4&J/4\\J^{\mathsf T}/4&I_3\end{pmatrix},\qquad\det\Sigma_0=\frac14,
  \label{eq:sm_site_covariance}
\end{equation}
where $J$ is the $4\times3$ matrix of ones.
The sum $\sum_jY_j$ has the mean $\frac L2(1,\ldots,1)$ and the covariance matrix $v\Sigma_0$.

The probability of $\sum_jY_j=\frac L2(1,\ldots,1)$ is the expectation value of the Kronecker delta of the integer vector $\sum_j(Y_j-\mathbb EY_j)$.
Writing this delta as $\int_{[-\pi,\pi]^7}d^7\theta\,e^{i\bm\theta\cdot\sum_j(Y_j-\mathbb EY_j)}/(2\pi)^7$ and using the independence of the $Y_j$, we obtain
\begin{equation}
  \Pr\Big[\sum_jY_j=\tfrac L2(1,\ldots,1)\Big]=\int_{[-\pi,\pi]^7}\frac{d^7\theta}{(2\pi)^7}\prod_{j=1}^{L}\varphi_j(\bm\theta),\qquad
  \varphi_j(\bm\theta)=\mathbb E\big[e^{i\bm\theta\cdot(Y_j-\mathbb EY_j)}\big].
  \label{eq:sm_fourier}
\end{equation}
Since $Y_j$ takes the value $\bm0$ and every unit vector is a difference of two values of $Y_j$, $\abs{\varphi_j(\bm\theta)}<1$ for $\bm\theta\neq\bm0$.
In the two states considered below, the number of sites with $p_j(1-p_j)$ of order one is of order $v$, so the integrand decreases exponentially in $v$ away from $\bm\theta=\bm0$ and only small $\bm\theta$ contributes.

For small $\bm\theta$, the logarithm of the integrand is the sum over $j$ of the cumulant expansions of $\ln\varphi_j$.
The first-order terms vanish, and the second-order terms add up to $-v\,\bm\theta^{\mathsf T}\Sigma_0\bm\theta/2$.
The odd-order terms cancel, because $p_{L+1-j}=1-p_j$ and Eq.~\eqref{eq:sm_v_flip} make $Y_{L+1-j}-\mathbb EY_{L+1-j}$ distributed as $-(Y_j-\mathbb EY_j)$.
We set $\bm\theta=\bm u/\sqrt v$ and expand $\sum_j\ln\varphi_j$ in $\bm u$,
\begin{equation}
  \sum_{j=1}^{L}\ln\varphi_j\Big(\frac{\bm u}{\sqrt v}\Big)=-\frac12\bm u^{\mathsf T}\Sigma_0\bm u+\frac{1}{4!\,v^2}\sum_{j=1}^{L}\sum_{a,b,c,d=1}^{7}\kappa^{(j)}_{abcd}u_au_bu_cu_d+O(v^{-2}),
\end{equation}
where $\kappa^{(j)}_{abcd}=\partial_{\theta_a}\partial_{\theta_b}\partial_{\theta_c}\partial_{\theta_d}\ln\varphi_j(\bm0)$ is the fourth cumulant of $Y_j$.
Taking the exponential and expanding it to first order in the second term gives
\begin{equation}
  \prod_{j=1}^{L}\varphi_j\Big(\frac{\bm u}{\sqrt v}\Big)=e^{-\bm u^{\mathsf T}\Sigma_0\bm u/2}\left[1+\frac{1}{4!\,v^2}\sum_{j=1}^{L}\sum_{a,b,c,d=1}^{7}\kappa^{(j)}_{abcd}u_au_bu_cu_d+O(v^{-2})\right].
\end{equation}
We change the integration variable to $\bm u$ with $d^7\theta=v^{-7/2}d^7u$ and extend the integration range to $\mathbb R^7$.
By the Gaussian integral, the term $1$ in the brackets gives $(2\pi v)^{-7/2}(\det\Sigma_0)^{-1/2}=2(2\pi v)^{-7/2}$.
By Wick's theorem, the average of $u_au_bu_cu_d$ with the weight $e^{-\bm u^{\mathsf T}\Sigma_0\bm u/2}$ is $(\Sigma_0^{-1})_{ab}(\Sigma_0^{-1})_{cd}+(\Sigma_0^{-1})_{ac}(\Sigma_0^{-1})_{bd}+(\Sigma_0^{-1})_{ad}(\Sigma_0^{-1})_{bc}$.
Since $\kappa^{(j)}_{abcd}$ is symmetric in its indices, the three terms contribute equally, and the fourth-order term gives the coefficient $3/4!=1/8$.
We thus obtain
\begin{equation}
  \Pr\Big[\sum_jY_j=\tfrac L2(1,\ldots,1)\Big]=\frac{2}{(2\pi v)^{7/2}}\left[1+\frac{1}{8v^2}\sum_{j=1}^{L}c_4(p_j)+O(v^{-2})\right],
  \label{eq:sm_edgeworth}
\end{equation}
where
\begin{equation}
  c_4(p_j)=\sum_{a,b,c,d=1}^{7}\kappa^{(j)}_{abcd}(\Sigma_0^{-1})_{ab}(\Sigma_0^{-1})_{cd}.
  \label{eq:sm_c4}
\end{equation}
This is the Edgeworth expansion~\cite{petrov1975sums}.
The same calculation for $N_\uparrow$ in the product state, whose single-site fourth cumulant is $p_j(1-p_j)[1-6p_j(1-p_j)]$, gives
\begin{equation}
  Z=\frac{1}{(2\pi v)^{1/2}}\left[1+\frac{1}{8v^2}\sum_{j=1}^{L}p_j(1-p_j)\big[1-6p_j(1-p_j)\big]+O(v^{-2})\right].
  \label{eq:sm_Z_expansion}
\end{equation}

To find $c_4(p)$ for a site with up-spin probability $p$, let $Q=(Y-\mathbb EY)^{\mathsf T}\Sigma_0^{-1}(Y-\mathbb EY)$ with $Y-\mathbb EY=(s_1-p,\ldots,s_4-p,t_1-p,t_2-p,t_3-p)$.
With the components $X_a$ of $Y-\mathbb EY$ and the fourth cumulant $\kappa_{abcd}$ of $Y$, the fourth moments are
\begin{equation}
  \mathbb E[X_aX_bX_cX_d]=\kappa_{abcd}+p^2(1-p)^2\left[(\Sigma_0)_{ab}(\Sigma_0)_{cd}+(\Sigma_0)_{ac}(\Sigma_0)_{bd}+(\Sigma_0)_{ad}(\Sigma_0)_{bc}\right].
\end{equation}
Multiplying both sides by $(\Sigma_0^{-1})_{ab}(\Sigma_0^{-1})_{cd}$ and summing over $a,b,c,d$, we find that the left-hand side becomes $\mathbb E[Q^2]$ and the term $\kappa_{abcd}$ on the right-hand side becomes $c_4(p)$ by Eq.~\eqref{eq:sm_c4}.
Since $\Sigma_0$ and $\Sigma_0^{-1}$ are $7\times7$ symmetric matrices, the first term in the brackets gives $[\Tr(\Sigma_0^{-1}\Sigma_0)]^2=49$, and the second and third terms each give $\Tr(\Sigma_0^{-1}\Sigma_0\Sigma_0^{-1}\Sigma_0)=7$.
Thus $c_4(p)=\mathbb E[Q^2]-63p^2(1-p)^2$.
The matrix $\Sigma_0^{-1}$ is
\begin{equation}
  \Sigma_0^{-1}=\begin{pmatrix}I_4+\frac34J_4&-J\\-J^{\mathsf T}&I_3+J_3\end{pmatrix},
  \label{eq:sm_sigma0_inverse}
\end{equation}
where $J_n$ is the $n\times n$ matrix of ones.
Substituting it and replacing $\sum_{i=1}^{3}(t_i-p)$ with $(\abs{s}-4p)-(t_4-p)$, we obtain
\begin{equation}
  Q=\sum_{i=1}^{4}\left[(s_i-p)^2+(t_i-p)^2\right]-\frac14(\abs{s}-4p)^2.
  \label{eq:sm_Q}
\end{equation}
Since $s_i$ and $t_i$ are $0$ or $1$, $(s_i-p)^2=(1-2p)(s_i-p)+p(1-p)$, and likewise for $t_i$.
Together with $\sum_i(s_i-p)=\sum_i(t_i-p)=\abs{s}-4p$, this gives
\begin{equation}
  Q=8p(1-p)+2(1-2p)(\abs{s}-4p)-\frac14(\abs{s}-4p)^2.
  \label{eq:sm_Q_closed}
\end{equation}
Thus $Q$ depends only on $\abs{s}$.
The number $\abs{s}$ follows the binomial distribution with four trials and probability $p$, and the second, third, and fourth moments of $\abs{s}-4p$ are $4p(1-p)$, $4p(1-p)(1-2p)$, and $4p(1-p)+24p^2(1-p)^2$.
These moments give $\mathbb E[Q^2]=\frac{49}{4}p(1-p)+\frac32p^2(1-p)^2$ and hence $c_4(p)=\frac{49}{4}p(1-p)-\frac{123}{2}p^2(1-p)^2$.
As in Eq.~\eqref{eq:sm_probability_Y}, we divide Eq.~\eqref{eq:sm_edgeworth} by the fourth power of Eq.~\eqref{eq:sm_Z_expansion}.
Using $c_4(p)-4p(1-p)[1-6p(1-p)]=\frac{33}{4}p(1-p)-\frac{75}{2}p^2(1-p)^2$ and Eq.~\eqref{eq:sm_variance}, we obtain
\begin{equation}
  \nu_2(\xi)=\frac{2}{(2\pi v)^{3/2}}\left[1+\frac{33}{32v}-\frac{75}{16v^2}\sum_{j=1}^{L}p_j^2(1-p_j)^2+O(v^{-2})\right].
  \label{eq:sm_nu2_expansion}
\end{equation}

\subsubsection{\label{subsec:sm_dicke_asym}Asymptotic form at \texorpdfstring{$\Delta=-1$}{Delta=-1}}

For $\ket{\mathrm D_L}$, all $p_j$ are $1/2$, so $v=L/4$ and $\sum_jp_j^2(1-p_j)^2=L/16$.
The correction in Eq.~\eqref{eq:sm_nu2_expansion} becomes $33/(8L)-75/(16L)=-9/(16L)$, and we obtain
\begin{equation}
  \nu_2(\mathrm D_L)=\frac{4\sqrt2}{\pi^{3/2}}L^{-3/2}\left[1-\frac{9}{16L}+O(L^{-2})\right].
  \label{eq:sm_dicke_purity_asymptotics}
\end{equation}
This yields Eq.~\eqref{eq:dicke_asymptotic} of the main text.

\subsubsection{\label{subsec:sm_kink}Equal-weight superposition of domain-wall states for \texorpdfstring{$\Delta<-1$}{Delta<-1}}

For $\Delta<-1$, we write $\Delta=-\cosh\eta$ with $\eta>0$ and add to $H$ the boundary fields of Refs.~\cite{alcaraz1995anisotropic,koma1998complete},
\begin{equation}
  H_+=H-\frac{\sinh\eta}{2}\left(S_L^z-S_1^z\right).
  \label{eq:sm_kink_hamiltonian}
\end{equation}
The rotated Hamiltonian is a sum of terms on neighboring sites,
\begin{equation}
  UH_+U^\dagger+\frac{(L-1)\cosh\eta}{4}=\sum_{j=1}^{L-1}\left[-S_j^xS_{j+1}^x-S_j^yS_{j+1}^y-\cosh\eta\left(S_j^zS_{j+1}^z-\frac14\right)-\frac{\sinh\eta}{2}\left(S_{j+1}^z-S_j^z\right)\right],
  \label{eq:sm_kink_bond}
\end{equation}
where the last terms in the brackets add up to the boundary fields.
The term on sites $j$ and $j+1$ has $\ket\uparrow_j\ket\uparrow_{j+1}$ and $\ket\downarrow_j\ket\downarrow_{j+1}$ as eigenstates with the eigenvalue $0$, and it is represented by the matrix $\frac12\begin{pmatrix}e^\eta&-1\\-1&e^{-\eta}\end{pmatrix}$ in the basis $\ket\uparrow_j\ket\downarrow_{j+1},\ket\downarrow_j\ket\uparrow_{j+1}$.
This matrix has the eigenvalues $0$ and $\cosh\eta$, and its eigenvector with the eigenvalue $0$ is $\ket\uparrow_j\ket\downarrow_{j+1}+e^\eta\ket\downarrow_j\ket\uparrow_{j+1}$.
Every term is thus positive semidefinite and annihilates any state whose amplitude grows by $e^\eta$ whenever an up spin moves one site to the right.
At $N_\uparrow=L/2$, such a state is
\begin{equation}
  \ket{\xi_+}\propto\exp\Big[\eta\sum_{j=1}^{L}\Big(j-\frac{L+1}{2}\Big)n_{j\uparrow}\Big]\ket{\mathrm D_L},
  \label{eq:sm_kink_state}
\end{equation}
because all configurations have the same amplitude in $\ket{\mathrm D_L}$ and moving an up spin one site to the right increases the sum in the exponent by one.
Since the boundary fields are diagonal in the $S^z$ basis, the argument of Sec.~\ref{sec:sm_ground_state} shows that $\ket{\xi_+}$ is the unique ground state of $UH_+U^\dagger$ at $N_\uparrow=L/2$.

The exponential in Eq.~\eqref{eq:sm_kink_state} is a product of factors of the individual sites and commutes with $N_\uparrow$.
The factor of site $j$ multiplies the amplitude of $\ket\uparrow$ relative to that of $\ket\downarrow$ by $e^{\eta(j-(L+1)/2)}$.
Since $\ket{\mathrm D_L}$ is the state of Eq.~\eqref{eq:sm_projected_state} with all $p_j=1/2$, $\ket{\xi_+}$ is the state of Eq.~\eqref{eq:sm_projected_state} with $\sqrt{p_j/(1-p_j)}=e^{\eta(j-(L+1)/2)}$, that is,
\begin{equation}
  p_j=\frac{1}{1+e^{-\eta(2j-L-1)}}.
  \label{eq:sm_kink_profile}
\end{equation}
In the product state, the spins are down near site $1$ and up near site $L$.
The variance $p_j(1-p_j)=1/[4\cosh^2(\eta(j-\frac{L+1}{2}))]$ of each spin is appreciable only within about $1/\eta$ sites of the center.
The state $\ket{\xi_-}=\mathcal R\ket{\xi_+}$ has the up-spin probabilities $1-p_j$.
As the state to be compared with the ground state of the chain without boundary fields, we use the $\mathcal R$-invariant superposition
\begin{equation}
  \ket{\xi_{\rm s}}=\frac{\ket{\xi_+}+\ket{\xi_-}}{\sqrt{2+2\braket{\xi_+}{\xi_-}}}.
  \label{eq:sm_kink_symmetrized}
\end{equation}

For $\eta\ll1$ and $L\eta\gg1$, the sums for $\ket{\xi_+}$ become integrals,
\begin{equation}
  v\simeq\int_{-\infty}^{\infty}\frac{dj}{4\cosh^2(\eta(j-\frac{L+1}{2}))}=\frac{1}{2\eta},\qquad
  \sum_jp_j^2(1-p_j)^2\simeq\int_{-\infty}^{\infty}\frac{dj}{16\cosh^4(\eta(j-\frac{L+1}{2}))}=\frac{1}{12\eta},
  \label{eq:sm_kink_sums}
\end{equation}
and the correction in Eq.~\eqref{eq:sm_nu2_expansion} becomes $\frac{33}{16}\eta-\frac{25}{16}\eta=\frac\eta2$.
We thus obtain
\begin{equation}
  \nu_2(\xi_+)=2\Big(\frac\eta\pi\Big)^{3/2}\left[1+\frac\eta2+O(\eta^2)\right].
\end{equation}

Finally, we find the purity of $\ket{\xi_{\rm s}}$.
Inserting $\ket{\xi_{\rm s}}$ into Eq.~\eqref{eq:sm_nu2_V} gives
\begin{equation}
  \nu_2(\xi_{\rm s})=\frac{1}{(2+2\braket{\xi_+}{\xi_-})^4}\sum_{\sigma,\tau\in\{+,-\}^4}\;\sum_{\bm s_1,\ldots,\bm s_4}\;\sum_{\bm t_1,\ldots,\bm t_4}\;\prod_{i=1}^{4}\overline{\xi_{\tau_i}(\bm t_i)}\,\xi_{\sigma_i}(\bm s_i)\prod_{j=1}^{L}V_{t^{(j)}s^{(j)}}.
\end{equation}
The term with all $\sigma_i,\tau_i$ equal to $+$ is $\nu_2(\xi_+)$, and the term with all of them equal to $-$ also equals $\nu_2(\xi_+)$ by $\xi_-(\bm s)=\xi_+(\bar{\bm s})$ and Eq.~\eqref{eq:sm_v_flip}.
Here $\bar{\bm s}$ is the configuration obtained from $\bm s$ by flipping all spins.

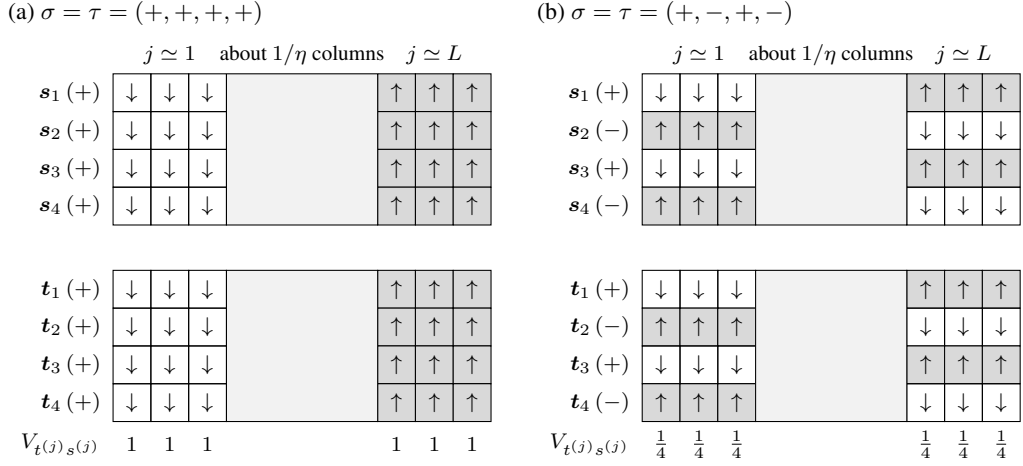
\begin{figure}[htbp]
  \centering
  \begin{tikzpicture}[x=5mm,y=5mm]
    \foreach \dx/\signs/\vval/\ttl in {0/{0,0,0,0,0,0,0,0}/1/{(a) $\sigma=\tau=(+,+,+,+)$},14/{0,1,0,1,0,1,0,1}/\tfrac14/{(b) $\sigma=\tau=(+,-,+,-)$}}{
      \begin{scope}[xshift=\dx*5mm]
        \foreach \sg [count=\i] in \signs{
          \pgfmathsetmacro{\yt}{\i<=4 ? 1-\i : -0.2-\i}
          \foreach \c in {0,1,2,7,8,9}{
            \pgfmathtruncatemacro{\up}{(\c<5) == (\sg==1)}
            \ifnum\up=1
              \fill[gray!30] (\c,\yt-1) rectangle (\c+1,\yt);
              \node[font=\footnotesize] at (\c+0.5,\yt-0.5) {$\uparrow$};
            \else
              \node[font=\footnotesize] at (\c+0.5,\yt-0.5) {$\downarrow$};
            \fi
            \draw (\c,\yt-1) rectangle (\c+1,\yt);
          }
          \pgfmathtruncatemacro{\k}{\i<=4 ? \i : \i-4}
          \ifnum\i<5
            \node[anchor=east,font=\footnotesize] at (-0.1,\yt-0.5) {$\bm s_{\k}\,(\ifnum\sg=0 +\else -\fi)$};
          \else
            \node[anchor=east,font=\footnotesize] at (-0.1,\yt-0.5) {$\bm t_{\k}\,(\ifnum\sg=0 +\else -\fi)$};
          \fi
        }
        \foreach \ytop in {0,-5.2}{
          \fill[gray!10] (3,\ytop-4) rectangle (7,\ytop);
          \draw (3,\ytop-4) rectangle (7,\ytop);
        }
        \foreach \c in {0,1,2,7,8,9}{
          \node[font=\footnotesize] at (\c+0.5,-9.8) {$\vval$};
        }
        \node[anchor=east,font=\footnotesize] at (-0.1,-9.8) {$V_{t^{(j)}s^{(j)}}$};
        \node[font=\footnotesize] at (1.5,0.5) {$j\simeq1$};
        \node[font=\footnotesize] at (5,0.5) {about $1/\eta$ columns};
        \node[font=\footnotesize] at (8.5,0.5) {$j\simeq L$};
        \node[anchor=west,font=\small] at (-3,1.6) {\ttl};
      \end{scope}
    }
  \end{tikzpicture}
  \caption{Examples of the arrays of spins in the $(\sigma,\tau)$ terms of $\nu_2(\xi_{\rm s})$.
    The upper four rows are $\bm s_1,\ldots,\bm s_4$, the lower four rows are $\bm t_1,\ldots,\bm t_4$, and the signs in parentheses are $\sigma_i$ and $\tau_i$.
    Up spins are shown in gray, and the spins in the central region of about $1/\eta$ columns are not drawn.
    The labels $j\simeq1$ and $j\simeq L$ indicate the columns with $\abs{j-\frac{L+1}{2}}\gg1/\eta$.
    In these columns, each spin is taken in the orientation with the larger amplitude, which is down at $j\simeq1$ and up at $j\simeq L$ in the rows from $\ket{\xi_+}$ and the reverse in the rows from $\ket{\xi_-}$.
    The bottom row shows $V_{t^{(j)}s^{(j)}}$ of column $j$, which is $1$ for all columns in (a) and $1/4$ for all columns in (b).}
  \label{fig:sm_kink_columns}
\end{figure}

Each remaining term contains rows $\bm s_i,\bm t_i$ from both $\ket{\xi_+}$ and $\ket{\xi_-}$.
At a site $j$ with $\frac{L+1}{2}-j\gg1/\eta$, the spins in the rows from $\ket{\xi_+}$ are mostly down and those in the rows from $\ket{\xi_-}$ are mostly up.
The eight spins of $s^{(j)}$ and $t^{(j)}$ in a remaining term are thus not all equal, and Eq.~\eqref{eq:sm_v_block} and the entries above it give $V_{t^{(j)}s^{(j)}}\leq1/4$, whereas $s^{(j)}=t^{(j)}={\downarrow\downarrow\downarrow\downarrow}$ and $V_{t^{(j)}s^{(j)}}=1$ in $\nu_2(\xi_+)$ (Fig.~\ref{fig:sm_kink_columns}).
An up spin in a row from $\ket{\xi_+}$ or a down spin in a row from $\ket{\xi_-}$ can make $V_{t^{(j)}s^{(j)}}=1$, but it multiplies the amplitude of its row by $\sqrt{p_j/(1-p_j)}\ll1/4$.
The same holds for $j-\frac{L+1}{2}\gg1/\eta$ with up and down interchanged and $p_j$ replaced with $1-p_j$.
Each of the $L-O(1/\eta)$ sites away from the center thus multiplies the remaining terms by at most about $1/4$ compared with $\nu_2(\xi_+)$, so the remaining terms are negligible for $L\eta\gg1$.

The product of the amplitudes of $\ket{\xi_+}$ and $\ket{\xi_-}$ for the same configuration $\bm s$ contains one factor $\sqrt{p_j}$ and one factor $\sqrt{1-p_j}$ at each site and is therefore independent of $\bm s$.
The overlap is thus $\braket{\xi_+}{\xi_-}=\binom{L}{L/2}Z^{-1}\prod_j\sqrt{p_j(1-p_j)}$.
Since $\binom{L}{L/2}\leq2^L$, $Z^{-1}$ is of order $v^{1/2}$ by Eq.~\eqref{eq:sm_Z_expansion}, and $2\sqrt{p_j(1-p_j)}=1/\cosh(\eta(j-\frac{L+1}{2}))$ is exponentially small for $\abs{j-\frac{L+1}{2}}\gg1/\eta$, the overlap is also negligible.
In the large-$L$ limit, we thus obtain
\begin{equation}
  \nu_2(\xi_{\rm s})=\frac{2\,\nu_2(\xi_+)}{2^4}=\frac{\nu_2(\xi_+)}{8},
  \label{eq:sm_kink_factor}
\end{equation}
which together with $\nu_2(\xi_+)$ yields Eq.~\eqref{eq:domain_wall_mre} of the main text.

Table~\ref{tab:sm_kink} compares Eq.~\eqref{eq:domain_wall_mre} with the MRE sampled from the ground state of the chain without boundary fields.
The deviation is $1.8$ standard errors at $(L,\Delta)=(64,-1.02)$ and $0.1$ standard errors at $(128,-1.005)$.

\begin{table}[H]
  \centering
  \caption{Equation~\eqref{eq:domain_wall_mre} of the main text compared with the MRE sampled from the ground state of the chain without boundary fields.
    The errors are one standard error.
    Both rows have $L\eta\simeq12.8$.}
  \label{tab:sm_kink}
  \begin{tabular}{lrr}
    \hline\hline
    $(L,\Delta)$ & Eq.~\eqref{eq:domain_wall_mre} & sampling\\
    \hline
    $(64,-1.02)$ & 5.420203 & $5.4036\pm0.0091$\\
    $(128,-1.005)$ & 6.507912 & $6.5071\pm0.0096$\\
    \hline\hline
  \end{tabular}
\end{table}

\section{\label{sec:sm_numerics}Numerical details}

\subsection{\label{subsec:sm_state_preparation}State preparation and checks for convergence}

\begin{figure}[htbp]
  \centering
  \includegraphics{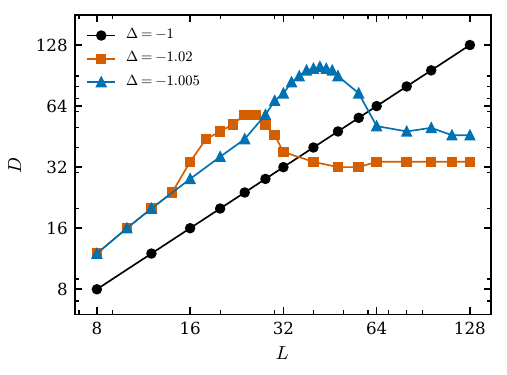}
  \caption{Bond dimension $D$ of the fermionic MPS used as the input of the sampling.
    At $\Delta=-1$, the exact Dicke MPS with $D=L$ is used without truncation.
    At $\Delta=-1.02$ and $-1.005$, the projected DMRG state with $\chi=256$ is compressed with $\varepsilon=10^{-10}$.}
  \label{fig:sm_bond_dimension}
\end{figure}

\begin{figure}[htbp]
  \centering
  \includegraphics[scale=0.75]{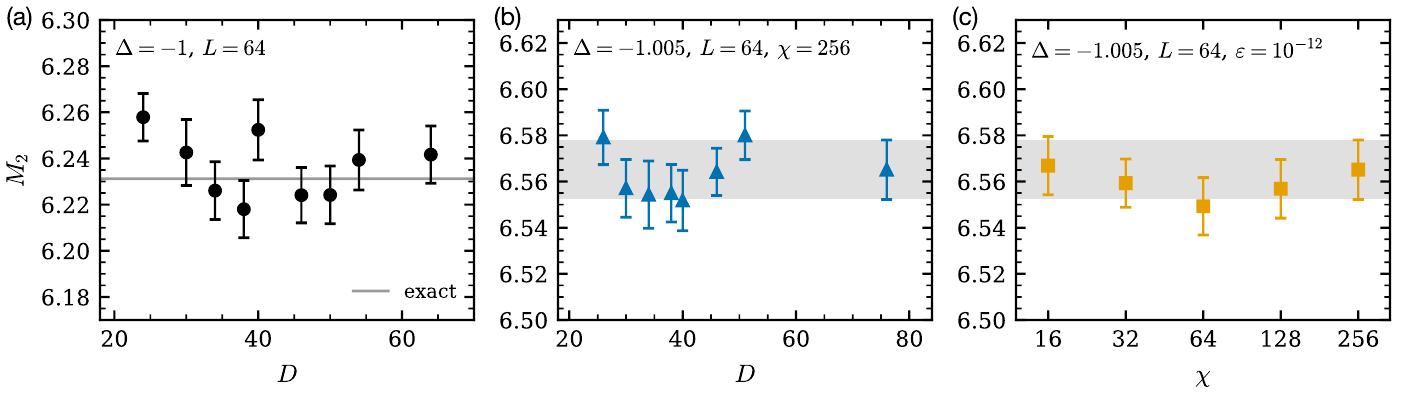}
  \caption{Dependence of the sampled MRE at $L=64$ on the approximations of the input state.
    (a) The exact Dicke MPS at $\Delta=-1$ compressed with the discarded weight $\varepsilon=10^{-2},\dots,10^{-12}$, plotted against the resulting bond dimension $D$, which increases as $\varepsilon$ decreases; the rightmost point is the uncompressed MPS and the line is the exact value.
    (b) The projected DMRG state at $\Delta=-1.005$ with $\chi=256$ compressed with the same values of $\varepsilon$ and plotted against $D$.
    The band is one standard error around $\widetilde M_2$ at $\varepsilon=10^{-12}$, the rightmost point.
    (c) The dependence on the DMRG bond dimension $\chi$ at $\Delta=-1.005$ with $\varepsilon=10^{-12}$; the band is the same as in (b).}
  \label{fig:sm_convergence}
\end{figure}

At $\Delta=-1$, the input of the sampling method is the exact MPS of the ground state over the $2L$ fermionic modes, with bond dimension $D=L$ and no truncation.
For $\Delta<-1$, we obtained the ground state at $N_\uparrow=N_\downarrow=L/2$ by two-site DMRG implemented in TeNPy~\cite{hauschild2018efficient,hauschild2024tensor} with bond dimension $\chi=256$, and projected the result onto the eigenspace of $\mathcal R$ with eigenvalue $(-1)^{L/2}$ given by Eq.~\eqref{eq:sm_spin_flip_eigenvalue}.
The projection is necessary because the two domain configurations are nearly degenerate on a finite chain and DMRG can converge to a symmetry-broken state.
The energy variance of the projected states is below $3\times10^{-12}$ for all $(L,\Delta)$, and the weight discarded by the DMRG truncation is below $10^{-15}$.
We then compressed the spin-chain MPS to the smallest bond dimension for which the discarded weight, i.e., the sum of the squared discarded Schmidt values over all bonds, does not exceed $\varepsilon=10^{-10}$, and converted it into an MPS over the $2L$ fermionic modes, which serves as the input of the sampling method.
Figure~\ref{fig:sm_bond_dimension} shows the resulting bond dimension $D$, which is largest near the maximum of $M_2$ and saturates at larger $L$.
At $\Delta=-1$, Fig.~\ref{fig:sm_convergence}(a) shows that $M_2$ at $L=64$ does not change beyond the statistical error when $\varepsilon$ is varied from $10^{-12}$ to $10^{-3}$ and deviates from the exact value only at $\varepsilon=10^{-2}$.
At $\Delta=-1.005$, Figs.~\ref{fig:sm_convergence}(b) and (c) show that $M_2$ at $L=64$ does not change beyond the statistical error when $\varepsilon$ is varied over the same range or $\chi$ from $16$ to $256$.
The value $\varepsilon=10^{-10}$ used for the data at $\Delta<-1$ is thus far below the discarded weight at which $M_2$ starts to change.

\subsection{\label{subsec:sm_cost}Sampling cost}

The runtime of the sampling is the time per sample multiplied by the number of samples $\mathcal N$ needed for a target standard error $\epsilon_M$ of $M_2$.
We discuss the two factors separately for the states in Fig.~\ref{fig:xxz_mre}(a), and quote $\mathcal N$ for $\epsilon_M=0.01$ throughout.

\subsubsection{\label{subsec:sm_time_per_sample}Time per sample}

\begin{figure}[htbp]
  \centering
  \includegraphics{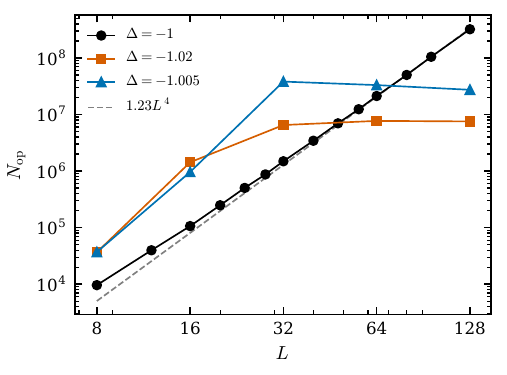}
  \caption{Number $N_{\rm op}$ of floating-point operations in the matrix products per sample for the states in Fig.~\ref{fig:xxz_mre}(a), at $L=8,16,32,64,128$ for $\Delta<-1$.
    The dashed line is $1.23L^4$, obtained by setting $n=2L$ and $D=L$ in $\mathcal O(nD^3)$, with the coefficient fitted to the points at $\Delta=-1$ and $L\geq64$.}
  \label{fig:sm_operations}
\end{figure}

Figure~\ref{fig:sm_operations} shows the number $N_{\rm op}$ of floating-point operations per sample, i.e., per evaluation of $h$, counted over all matrix products in our implementation.
Since the exact MPS at $\Delta=-1$ has $D=L$, setting $n=2L$ and $D=L$ in $\mathcal O(nD^3)$ gives $N_{\rm op}\propto L^4$ at large $L$.
The curve $1.23L^4$, fitted to the data for $L\geq64$, agrees with them within $3\%$.
At $\Delta<-1$, $N_{\rm op}$ stops growing once $D$ saturates, because the bond dimension is large only near the domain wall and remains small in the polarized regions.
Even for the exact MPS at $\Delta=-1$ and $L=128$, the most expensive case, one sample takes $24$~ms on average on one core of an Intel Xeon Gold 5318Y, so that the $1.4\times10^7$ samples needed for $\epsilon_M=0.01$ amount to a total CPU time of about $93$ hours (note that this sampling can be parallelized on separate cores).

\subsubsection{\label{subsec:sm_samples}Number of samples}

\begin{figure}[htbp]
  \centering
  \includegraphics[scale=0.75]{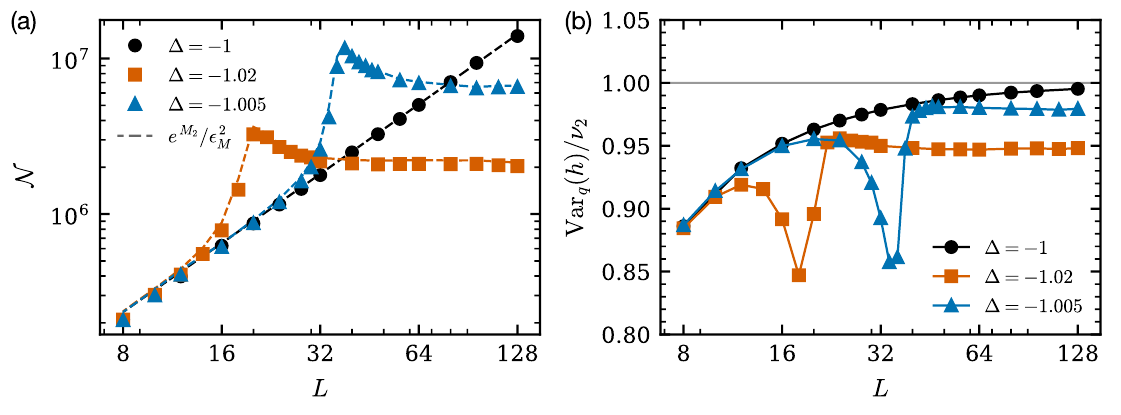}
  \caption{Number of samples for the states in Fig.~\ref{fig:xxz_mre}(a) in the main text.
    (a) The number of samples $\mathcal N=\operatorname{Var}_q(h)/(\epsilon_M^2\nu_2^2)$ needed for $\epsilon_M=0.01$, evaluated from the sample variance of $h$ and the sampled purity of each state.
    The dashed lines show $e^{M_2}/\epsilon_M^2$, with the exact $M_2$ at $\Delta=-1$ and the sampled $M_2$ at $\Delta<-1$.
    (b) The ratio of the sample variance of $h$ to the purity.}
  \label{fig:sm_samples}
\end{figure}

The number of samples needed for $\epsilon_M$ is $\mathcal N=\operatorname{Var}_q(h)/(\epsilon_M^2\nu_2^2)$.
Figure~\ref{fig:sm_samples}(a) shows that $\mathcal N\simeq e^{M_2}/\epsilon_M^2$ is determined by $M_2$ alone for all values of $\Delta$.
This follows from the fact that the ratio $\operatorname{Var}_q(h)/\nu_2$ lies between $0.85$ and $1$ as shown in Fig.~\ref{fig:sm_samples}(b).
At $\Delta=-1$, the $\mathcal N$ evaluated from the samples grows as $L^{3/2}$ and reaches $1.4\times10^7$ at $L=128$.
At $\Delta<-1$, $\mathcal N$ saturates together with $M_2$, at about $2.1\times10^6$ for $\Delta=-1.02$ and $6.5\times10^6$ for $\Delta=-1.005$.

\section{\label{sec:sm_extensive}Number of samples for an extensive MRE}

In this section, we examine the number of samples $\mathcal N$ for ground states whose MRE is extensive, using a chain of spinless fermions.
We take a chain of $L$ sites with one fermionic mode $c_j$ per site, i.e., $n=L$ modes, and the Hamiltonian
\begin{equation}
  H_{tV}=-t\sum_{j=1}^{L}\left(c_{j+1}^\dagger c_j+c_j^\dagger c_{j+1}\right)+V\sum_{j=1}^{L}n_jn_{j+1},
  \label{eq:sm_tv_chain}
\end{equation}
where $n_j=c_j^\dagger c_j$.
We consider the ground state at half filling, i.e., with $L/2$ fermions for even $L$, and impose the boundary condition $c_{L+1}=(-1)^{L/2+1}c_1$.
We parametrize $V$ by the Luttinger parameter $K$ of the ground state, $V=-2t\cos[\pi/(2K)]$ for $-2t<V\leq2t$~\cite{giamarchi2003quantum}, and use $K=2,3,4,5$.
By the Jordan-Wigner transformation followed by a rotation of the spins on the even sites by $\pi$ about the $z$ axis, $H_{tV}$ with this boundary condition corresponds to the XXZ chain with periodic boundary conditions and anisotropy $V/(2t)$, which ranges from $-0.71$ to $-0.95$ for $K=2$ to $5$.
The Jordan-Wigner transformation represents each spin by one fermionic mode, whereas the main text represents it by the two modes $c_{j\uparrow},c_{j\downarrow}$ with one fermion per site, and the MRE of a spin state depends on this choice.
For example, the ground state at $V=0$ is Gaussian in the present representation, whereas in the representation of the main text it satisfies Eq.~\eqref{eq:sm_covariance} and is thus non-Gaussian.

We obtained the ground state by two-site DMRG with particle-number conservation implemented in TeNPy~\cite{hauschild2018efficient,hauschild2024tensor} with bond dimension up to $\chi=768$.
We then compressed the MPS to the smallest bond dimension for which the discarded weight defined in Sec.~\ref{subsec:sm_state_preparation} does not exceed $\varepsilon=10^{-7}$.
The total weight discarded by the DMRG truncation and the compression is below $3\times10^{-6}$ for all states.
We quote $\mathcal N$ for $\epsilon_M=0.01$, evaluated as in Fig.~\ref{fig:sm_samples}(a).

\begin{figure}[htbp]
  \centering
  \includegraphics[scale=0.75]{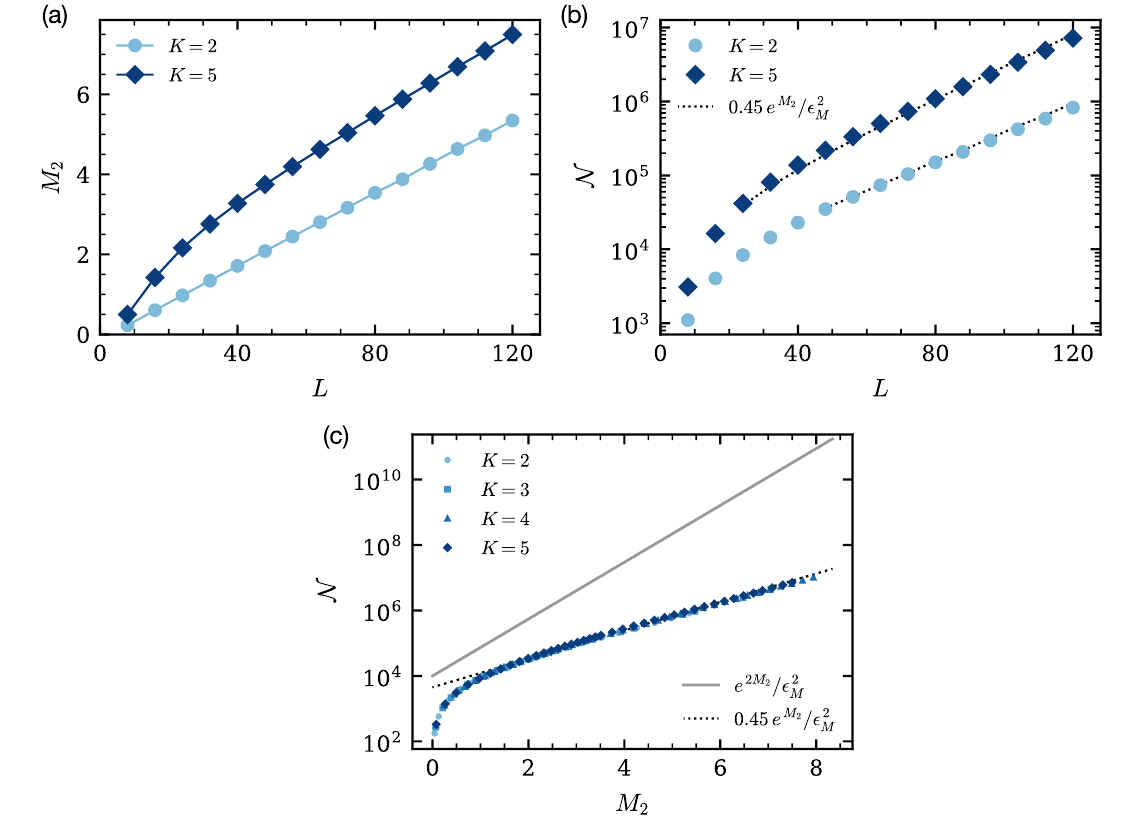}
  \caption{Number of samples for the ground state of Eq.~\eqref{eq:sm_tv_chain} at half filling.
    (a) Sampled MRE as a function of $L$ for the Luttinger parameters $K=2$ and $5$, connected by solid curves as guides.
    The standard errors are smaller than $0.01$ for all points.
    (b) The number of samples $\mathcal N$ needed for $\epsilon_M=0.01$ for the states in (a).
    The dotted lines show $0.45\,e^{M_2}/\epsilon_M^2$ with the sampled $M_2$ at the points with $M_2\geq2$.
    (c) $\mathcal N$ against the sampled $M_2$ for $K=2,3,4,5$ and $4\leq L\leq120$ ($L\leq96$ for $K=3$; $133$ points).
    The solid line is the upper bound $e^{2M_2}/\epsilon_M^2$, and the dotted line is $0.45\,e^{M_2}/\epsilon_M^2$.}
  \label{fig:sm_spinless_samples}
\end{figure}

As in Sec.~\ref{subsec:sm_samples}, $\mathcal N$ scales as $e^{M_2}$ rather than as the upper bound $e^{2M_2}$, which implies that the practical sampling cost is much lower than the worst-case bound.
Figure~\ref{fig:sm_spinless_samples}(a) shows that $M_2$ grows linearly with $L$ at large $L$ for $K=2$ and $5$, and reaches $7.5$ at $L=120$ for $K=5$.
Figure~\ref{fig:sm_spinless_samples}(b) shows that $\mathcal N$ grows exponentially with $L$ at large $L$ and reaches $8.3\times10^5$ for $K=2$ and $7.2\times10^6$ for $K=5$ at $L=120$.
Figure~\ref{fig:sm_spinless_samples}(c) shows that the points for $K=2,3,4,5$ lie on a single curve close to $0.45\,e^{M_2}/\epsilon_M^2$ for $M_2\gtrsim2$.
The largest $\mathcal N$ among these points is $1.03\times10^7$, for $K=4$ at $L=120$.
Here $0.45$ is the geometric mean of $\operatorname{Var}_q(h)/\nu_2$ over the $82$ points with $M_2\geq2$, where $\operatorname{Var}_q(h)$ and $\nu_2$ are evaluated from the samples.

\end{document}